\documentclass[journal]{IEEEtran}
\IEEEoverridecommandlockouts
\usepackage{cite}
\usepackage{amsmath,amssymb,amsfonts}
\usepackage{algorithmic}
\usepackage{graphicx}
\usepackage{textcomp}
\usepackage{xcolor}
\usepackage{float}
\usepackage{array}
\usepackage{booktabs}
\usepackage{amsmath}
\usepackage{multirow}
\usepackage{subcaption}

\usepackage{mathtools}
\usepackage{adjustbox}
\usepackage{tabulary}
\usepackage{colortbl}
\usepackage{array}
\newcolumntype{H}{>{\setbox0=\hbox\bgroup}c<{\egroup}@{}}

\def\BibTeX{{\rm B\kern-.05em{\sc i\kern-.025em b}\kern-.08em
    T\kern-.1667em\lower.7ex\hbox{E}\kern-.125emX}}
\begin{document}
\bstctlcite{IEEEexample:BSTcontrol}
\title{A Low-Power Streaming Speech Enhancement Accelerator For Edge Devices \\
}

\author{\IEEEauthorblockN{Ci-Hao Wu, and Tian-Sheuan Chang, \textit{Senior Member, IEEE}}
\thanks{This work was supported by the National Science and Technology Council, Taiwan, under Grant 111-2622-8-A49-018-SB, 110-2221-E-A49-148-MY3, and 110-2218-E-A49-015-MBK. The authors are with the Institute of Electronics, National Yang Ming Chiao Tung University, Taiwan. (e-mail: wucihou.ee09g@nctu.edu.tw
, tschang@nycu.edu.tw) }%
\thanks{Manuscript received XXXX XX, 2023; revised XXXX XX, XXXX.}
}
\maketitle

\begin{abstract}%

Transformer-based speech enhancement models yield impressive results. However, their heterogeneous and complex structure restricts model compression potential, resulting in greater complexity and reduced hardware efficiency. Additionally, these models are not tailored for streaming and low-power applications. Addressing these challenges, this paper proposes a low-power streaming speech enhancement accelerator through model and hardware optimization. The proposed high performance model is optimized for hardware execution with the co-design of model compression and target application, which reduces 93.9\% of model size by the proposed domain-aware and streaming-aware pruning techniques. The required latency is further reduced with batch normalization-based transformers. Additionally, we employed softmax-free attention, complemented by an extra batch normalization, facilitating simpler hardware design. The tailored hardware accommodates these diverse computing patterns by breaking them down into element-wise multiplication and accumulation (MAC). This is achieved through a 1-D processing array, utilizing configurable SRAM addressing, thereby minimizing hardware complexities and simplifying zero skipping. Using the TSMC 40nm CMOS process, the final implementation requires merely 207.8K gates and 53.75KB SRAM. It consumes only 8.08 mW for real-time inference at a 62.5MHz frequency.

\end{abstract}

\maketitle

\section{Introduction}

\begin{figure}[htbp]
\centering
\includegraphics[height=!,width=1.0\linewidth,keepaspectratio=true]{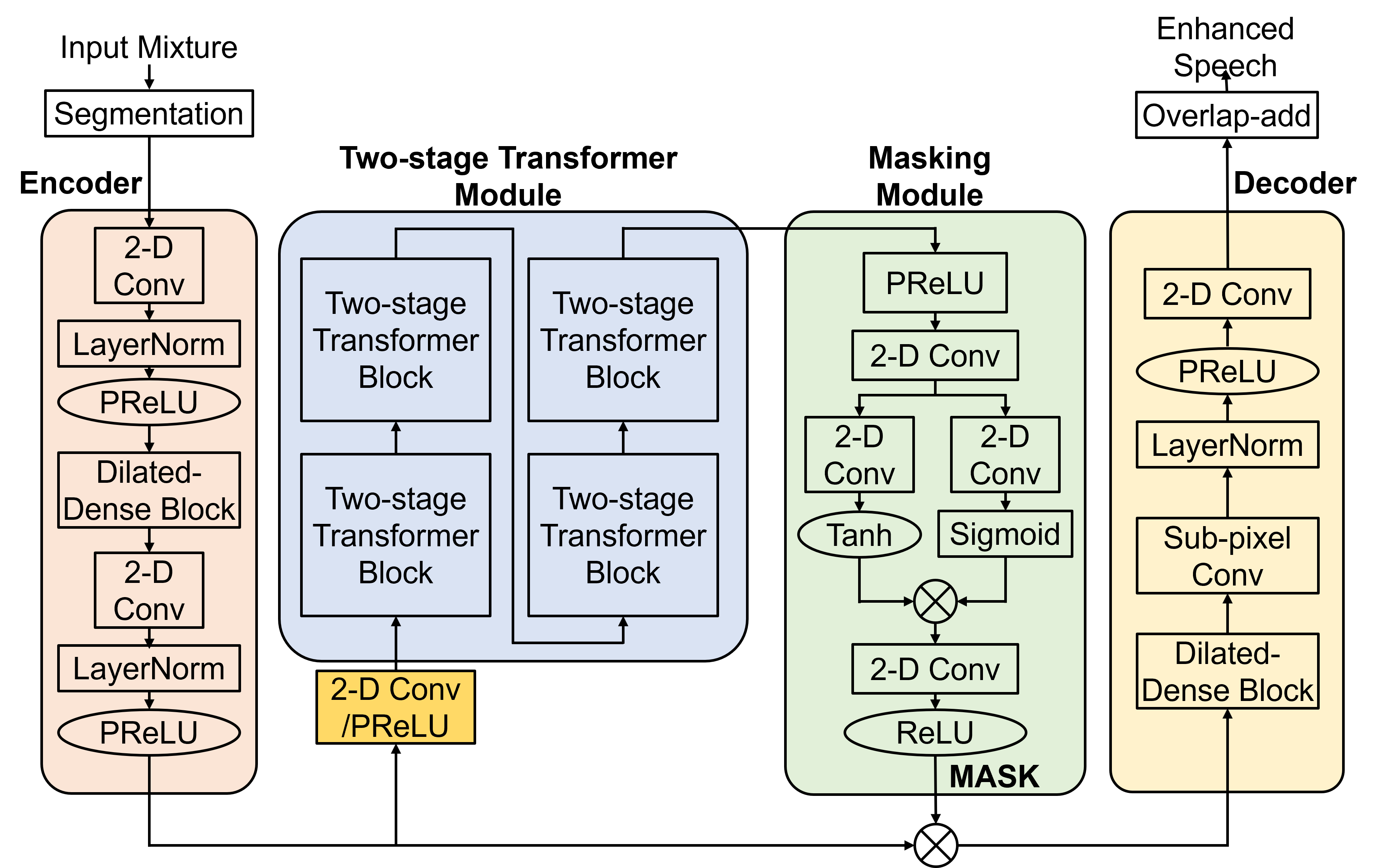}
\includegraphics[height=!,width=1.0\linewidth,keepaspectratio=true]{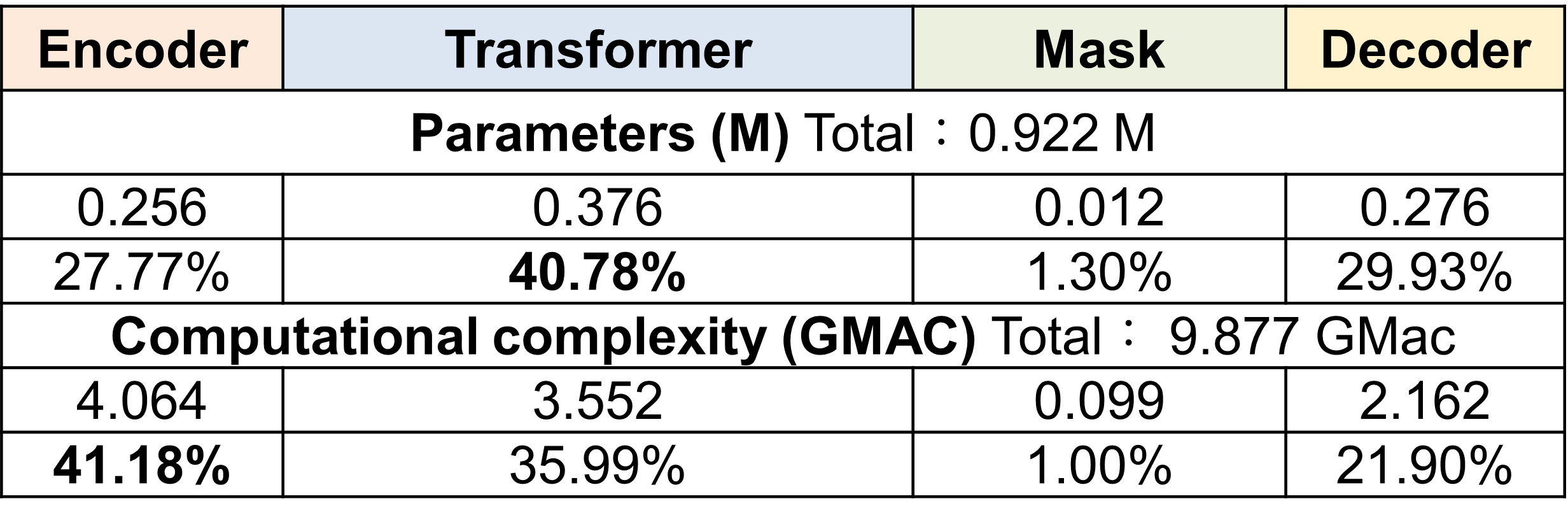}
\caption{The parameters and complexity distribution of the two-stage transformer neural network (TSTNN) \cite{TSTNN_2021}. The computational complexity is calculated with 8K samples per second.}
\label{analysisTSTNN}
\end{figure}

Deep learning-based speech enhancement (SE) surpasses traditional methods in enhancing speech intelligibility and quality. This enhancement is crucial for various natural language processing (NLP) tasks, including speech recognition, machine translation, and hearing aids. Transformer-based speech enhancement models, such as \cite{TSTNN_2021, DBT_Net}, have received significant attention in recent years due to their superior performance and parallel computing capabilities relative to other methods. The model shown in Fig.~\ref{analysisTSTNN} is heterogeneous, comprising an encoder and decoder that use convolutional neural networks (CNN) for speech extraction and restoration. In addition, it employs a masking module with transformers to filter out noise. However, its large model size and computational complexity become bottlenecks for low-power and real-time edge applications. Moreover, these models aren't optimized for streaming applications.

Addressing the needs for real-time and low-power solutions, this paper proposes a low-power design, achieved through model and architecture co-optimization. For hardware oriented model compression, we employ co-design of model compression and target application. We use \textit{domain-aware and streaming-aware pruning}, selectively trimming parts based on their significance to the SE task and streaming requirements. This method, combined with \textit{cross-domain masking and loss}, achieves a reduction of 93.9\% in model size and 94.9\% in complexity, all while maintaining comparable performance. We further decreased the model's latency by adopting constant batch normalization (BN) over dynamic layer normalization (LN)-based transformers—resulting in a 66\% cycle savings—and by implementing \textit{softmax-free attention with an additional BN}, which yields a 16X attention speedup. Our hardware design accommodates these diverse computing patterns by breaking them down into element-wise MACs. This is achieved using a \textit{1-D processing array with configurable SRAM addressing}, which minimizes hardware complexities, aligns with 1-D speech streaming processing, and facilitates straightforward data gating for zero skipping. Through these optimizations, we achieved reductions of 93.9\% in model size and 95\% in complexity. Moreover, our hardware design demands a mere 207.8K gate count and consumes 8.08 mW, making it apt for real-time streaming speech enhancement.

The structure of the paper is outlined as follows. Section II delves into related work concerning speech enhancement and deep learning accelerators. Section III presents our proposed optimizations for model complexity and hardware design. Section IV details our proposed hardware design. Section V showcases the experimental results. Section VI provides the conclusion.

\section{Related Works}
\subsection{Speech Enhancement}
Speech enhancement models primarily fall into two categories: time domain and time-frequency (T-F) domain. Time-domain techniques, such as those in \cite{Wavenet,DCCRN}, aim to predict a clean waveform directly from the noisy input. Methods in the T-F domain, as presented in \cite{TF_MASK_2018, Deepfilternet}, begin by transforming the noisy waveform into a spectrogram via the short-time Fourier transform (STFT). They then predict a mask for this spectrogram and subsequently reconstruct an enhanced waveform using the inverse STFT. A notable subset of these models, like those in \cite{DEMUCS,DCCRN}, employ architectures such as recurrent neural networks (RNN) or long-short-term memory (LSTM) to leverage temporal context information.

A surge in recent research, as highlighted in \cite{TSTNN_2021, DBT_Net, SETransformer}, has adopted the transformer architecture \cite{Attention} for its superior performance and capability for parallel computation. These studies integrate LSTM or Gate Recurrent Unit (GRU) modules with transformers to enhance positional information learning. Such an approach has shown potential to surpass the performance of models relying solely on RNN, LSTM, or even generative adversarial networks (GANs). For a comprehensive survey on the topic, readers can refer to \cite{das2021fundamentals}.

\subsection{Model Compression}
Model compression techniques, as discussed in \cite{deng2020model}, encompass pruning, quantization, and the introduction of simpler structures to reduce both the model size and complexity. Prior research on speech enhancement, such as \cite{tan2021towards}, has employed these general compression strategies specifically for CNN/LSTM-based models. However, these methods are not directly applicable to transformer-based models due to their heterogeneous layers. Additionally, these general methods often overlook potential application-specific compression opportunities.

\subsection{Deep Learning Accelerators}

In recent years, there's been a growing interest in deep learning accelerators (DLAs) for real-time model inference. While the majority of DLAs are tailored for vision tasks, characterized by numerous processing elements (PEs) and high memory bandwidth to accommodate substantial neural network computations, speech enhancement models present unique challenges. These include features like 1-D input, GRU structures, and transformers. For instance, many current DLAs lack support for GRUs owing to their intricate data flow. For speech applications, accelerators often favor RNN\cite{Chipmunk} or LSTM \cite{JSSC_2020} architectures, given their superior performance over CNNs in this domain. Additionally, \cite{NTUADSP} introduces an acoustic DSP integrated with a simplified CNN model, designed specifically for hearing assistive devices.

In the realm of transformer acceleration, \cite{ISCA2021_ELSA, HPCA2020_A3, acceltran, vitcod} introduce accelerators specifically designed to expedite the dot product operations within the transformer's self-attention mechanism. Additionally, \cite{hardware_acc_trans} unveils a specialized hardware accelerator that encompasses the entire transformer module. This design utilizes a systolic array for swift self-attention computation and extends native support for both LN and softmax operations. On another note, \cite{ISSCC2022_28nm} put forth a transformer processor designed to bypass weakly related tokens, targeting enhanced energy efficiency. However, this approach introduces an irregular and intricate computing structure. In summary, these transformer designs focus only on optimizing the attention part instead of the whole models, which is not sufficient for speech enhancement applications.

While the aforementioned designs predominantly cater to NLP and vision applications, our research pivots towards speech enhancement. This application demands streaming processing for real-time needs. The speech signals also has its distinct data distribution, when compared to NLP and vision. By capitalizing on these unique data characteristics, we've refined our approach through a synergistic hardware-software codesign strategy. It's worth noting that transformer-based speech enhancement models also necessitate CNN acceleration, a facet often overlooked in prior designs. Striking a balance in supporting both paradigms poses a challenge in maximizing hardware utilization.

\section{Hardware Oriented Model Optimization}
\label{chapter:software}

\subsection{Analysis and Design Challenges of TSTNN}

In this paper, we adopt TSTNN~\cite{TSTNN_2021}, a state-of-the-art transformer-based neural network, as our baseline owing to its commendable performance. Fig.~\ref{analysisTSTNN} illustrates the model architecture, highlighting the parameter and complexity distribution of TSTNN. In terms of parameter distribution, transformers account for approximately 40\%, whereas the encoder and decoder sections together contribute to nearly 57\%. Notably, the dense dilated block encompasses 53.36\% of the overall parameters, surpassing even the transformer components. The total model parameters and computational complexity are 0.922M and 9.877 GMACs, respectively, presenting challenges for deployment on milliwatt-scale edge devices in real-time scenarios. 

Aside from its considerable size and significant computational demands, TSTNN presents multiple structural challenges when it comes to streaming deployment. Primarily, the model doesn't operate as a causal system tailored for streaming, primarily due to the full-band attention within its transformer block. Moreover, the model features two blocks with high data dependencies: LN and softmax. These blocks hinder parallel execution, inevitably leading to increased latency. Furthermore, the model showcases a gamut of distinct computation patterns, including convolution, attention, GRU, and dilated convolution. Such diversity in computation patterns not only complicates hardware design but also introduces overhead, leading to diminished hardware efficiency.

\subsection{Overview Of The Proposed Approach}

In our quest to satisfy real-time requirements and achieve power efficiency without compromising performance, we tackle the aforementioned challenges by proposing a two-fold strategy: compressing the model using \textit{domain-aware and streaming-aware pruning} and adopting a hardware-friendly model design. Simultaneously, we ensure robust performance through the application of \textit{cross-domain masking and loss}.

In the context of domain-aware pruning, most contemporary pruning techniques, as discussed in \cite{deng2020model}, are tailored for homogeneous architectures. Given the distinct convolution/transformer structures present in our context, these methods are not directly applicable. As an alternative, our proposed pruning techniques focus on compressing the model based on two key criteria: the components' sensitivity to the speech enhancement task, termed as \textit{SE-aware pruning}, and the prerequisites of streaming inference, termed as \textit{streaming-aware pruning}. Using domain-aware pruning streamlines the identification of potential pruning regions within the model. Further details of this approach will be elaborated upon in the subsequent sections.

\subsection{Cross Domain Masking And Loss}
Deploying TSTNN on low-power edge devices necessitates substantial model compression, a process which often risks significant performance degradation. To counteract this, a viable strategy involves training a more robust model that delivers enhanced performance without expanding its size. While TSTNN employs masking in the time-domain and calculates loss across both spectrum- and time-domains for optimal performance, it overlooks frequency-related information during the masking process. The work in  \cite{Joint_TF_T} adopts the time-frequency masking but uses the frequency loss only. This paper, however, proposes using both time-frequency masking and time-frequency loss.  Consequently, speech inputs undergo initial processing via a STFT, serving as the spectrum input. Our proposed model, called the Time-Frequency Transformer Neural Network (TFTNN), takes advantage of this cross-domain approach in both its masking and loss calculations, aiming to boost performance. Consequently, our discussions in the following sections will center on TFTNN.

\subsection{Domain-aware Pruning}

Many existing pruning techniques, as highlighted in \cite{deng2020model}, are designed for homogeneous architectures. Directly applying them to our context is impractical, given the unique heterogeneous nature combining convolutional and transformer structures. To address this, we introduce domain-aware pruning methods that tailor compression based on the component's relevance to the speech enhancement task, termed as \textit{SE-aware pruning}, ensuring minimal performance degradation. Our domain-aware pruning approach encompasses four primary modifications: dilated residual block with channel splitting, pruning of the transformer block, mask module without gating mechanism, and replacing the parametric rectified linear unit (PReLU) with the rectified linear unit (ReLU).

\begin{figure}[htbp]
\begin{subfigure}{0.5\linewidth}
  \centering
 \includegraphics[height=!,width=\linewidth,keepaspectratio=true]{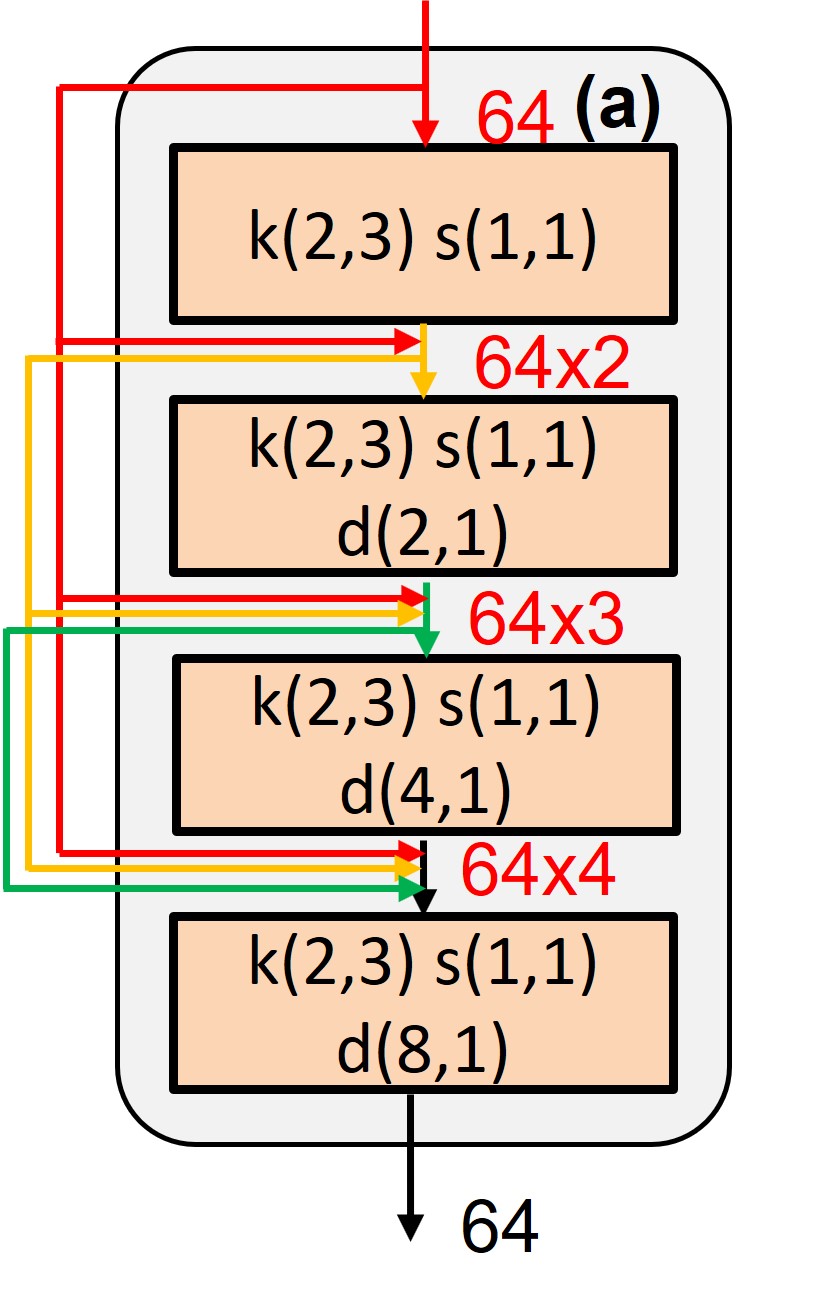}
\end{subfigure}
\begin{subfigure}{0.45\linewidth}
  \centering
 \includegraphics[height=!,width=\linewidth,keepaspectratio=true]{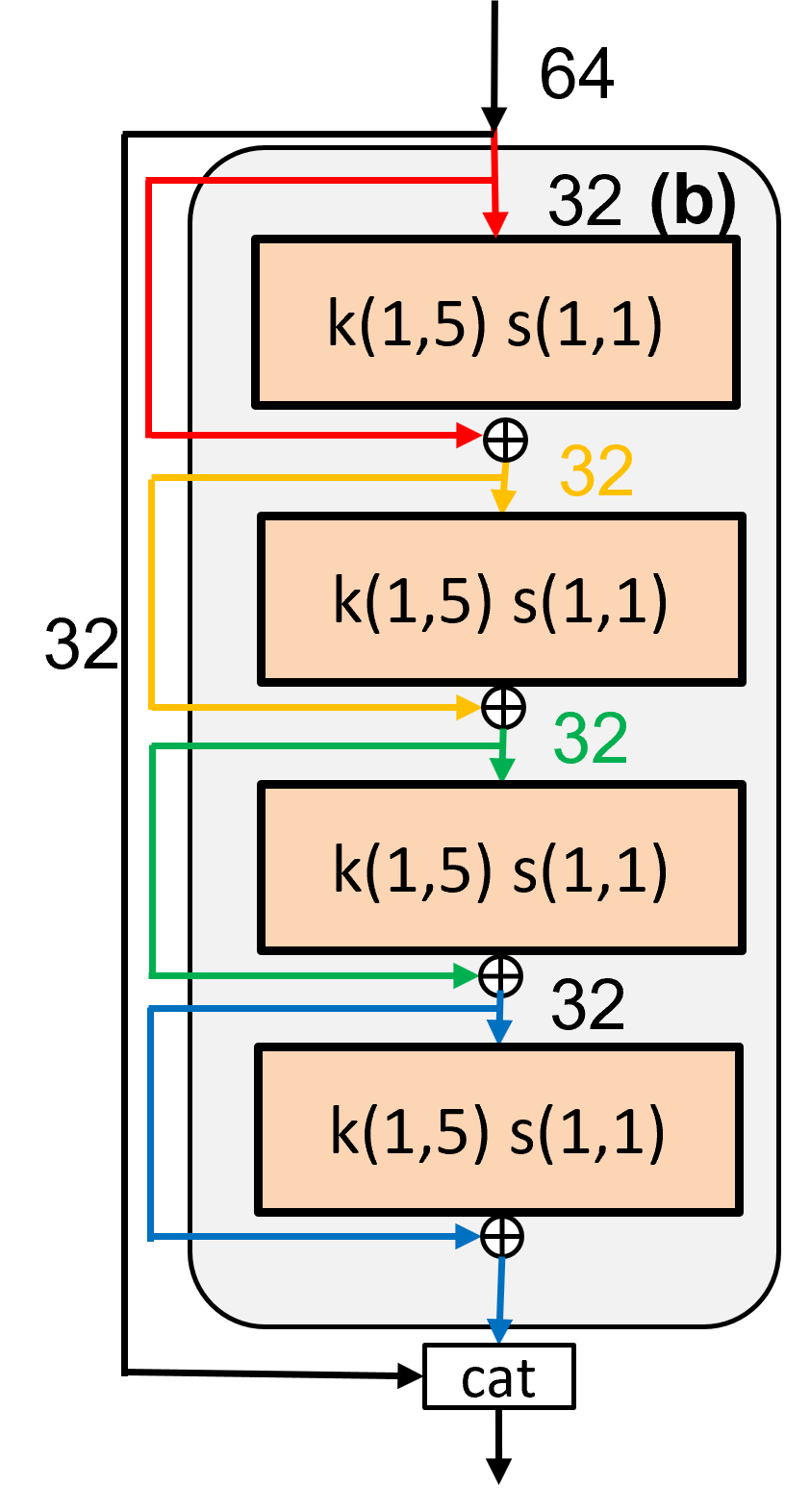}
 \end{subfigure}
\caption{(a) Dilated dense Block, and (b) dilated residual block with channel splitting. Each block is a convolution block with kernel size \textit{k}, stride \textit{s} and dilation rate \textit{d}. Each convolution is followed by LN/PReLU in the dilated dense block, and LN/ReLU in the dilated residual block.}
\label{csp_res_block}
\end{figure}

In the encoder/decoder segment, as depicted in Fig.\ref{csp_res_block}(a), the dense dilated block incorporates dense connections. This design choice increases the number of channels, subsequently amplifying both computational complexity and memory access demands. Notably, while dilation effectively captures long-range dependencies in speech signals, dense connections are not conducive to hardware implementations. To address this, we advocate for the dilated residual block illustrated in Fig.\ref{csp_res_block} (b). This block substitutes dense connections with more scalable residual ones, mitigating channel expansion. Additionally, by employing channel splitting, which processes only half the channels and bypsses half the channels, we achieve substantial reductions in complexity with almost the same performance. These modifications lead to a marked reduction of 90.2\% in parameters and a 90.0\% drop in GMACs for this specific block.

\begin{figure}[htbp]
\centering
\includegraphics[height=!,width=1\linewidth,keepaspectratio=true]{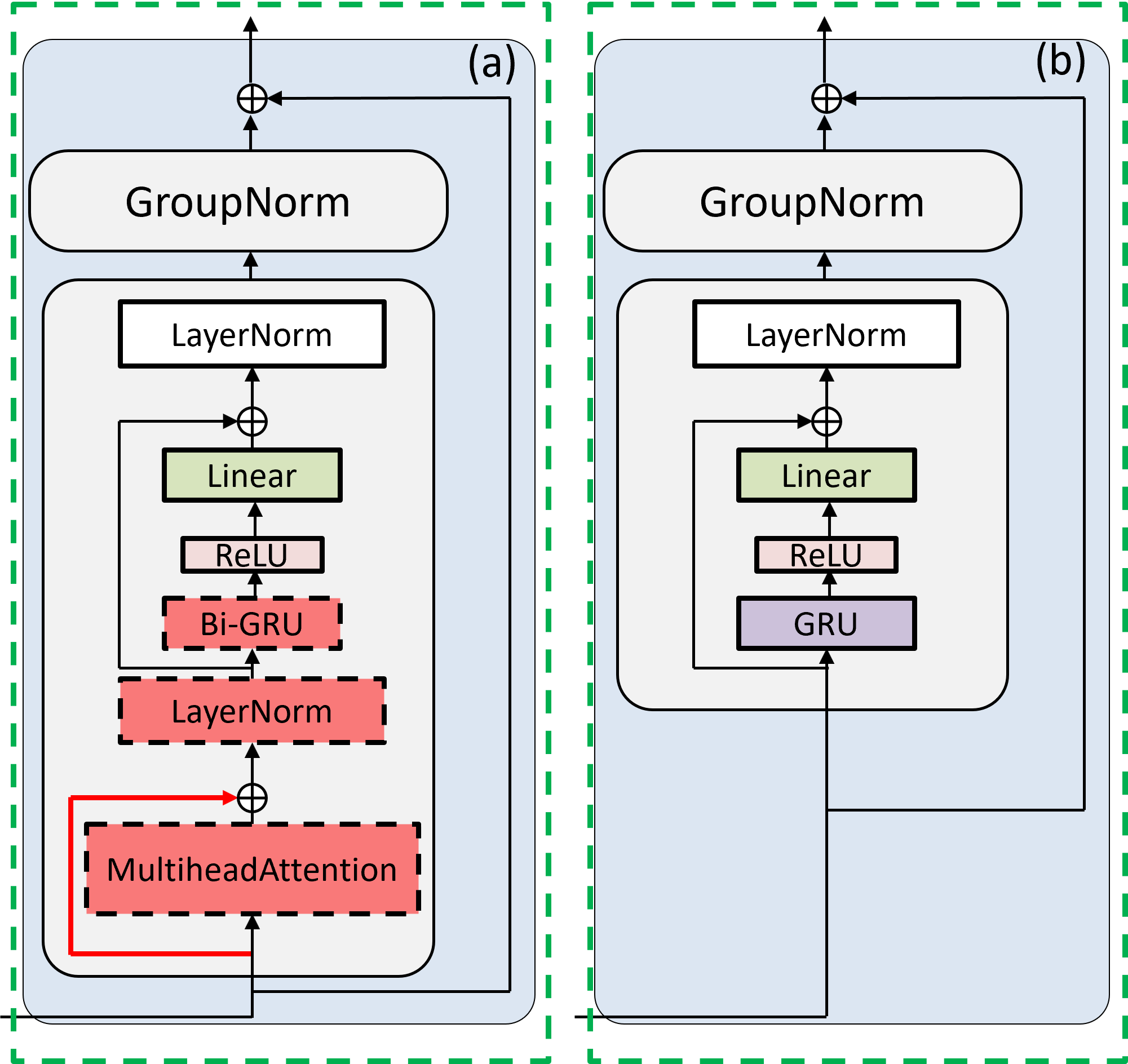}
\caption{Transformer (a) with full-band multi-head attention, and (b) without full-band multi-head attention.}
\label{modify_trans_block}
\end{figure}

\label{section:pruning for the transformer blocks}
The transformer block, illustrated in Fig.~\ref{modify_trans_block}(a), encompasses components like multi-head Attention (MHA), GRU, and linear modules. To reduce the number of parameters, we strategically reduce half of the channels within the GRU and MHA components. Our rationale for this adjustment is summarized below. The input channel corresponds to the \textit{embedding} dimension in MHA and the \textit{hidden} dimension in GRU. The \textit{embedding} and \textit{hidden} dimensions exert a lesser influence on the SE task's performance as compared to the \textit{length} dimension in both MHA and GRU based on our simulation.

Building on the channel pruning executed in the transformer blocks, we further reduce half of the channels in both the encoder and decoder segments, ensuring uniformity in channel count across the model. Given this halving of the model's channel count, retaining a deep network structure will improve performance slightly. Consequently, we streamline the architecture by cutting the number of transformer blocks from four down to two, a modification that, as our simulations indicate, retains comparable performance levels.

\begin{figure}[htbp]
\centering
\includegraphics[height=!,width=1.0\linewidth,keepaspectratio=true]{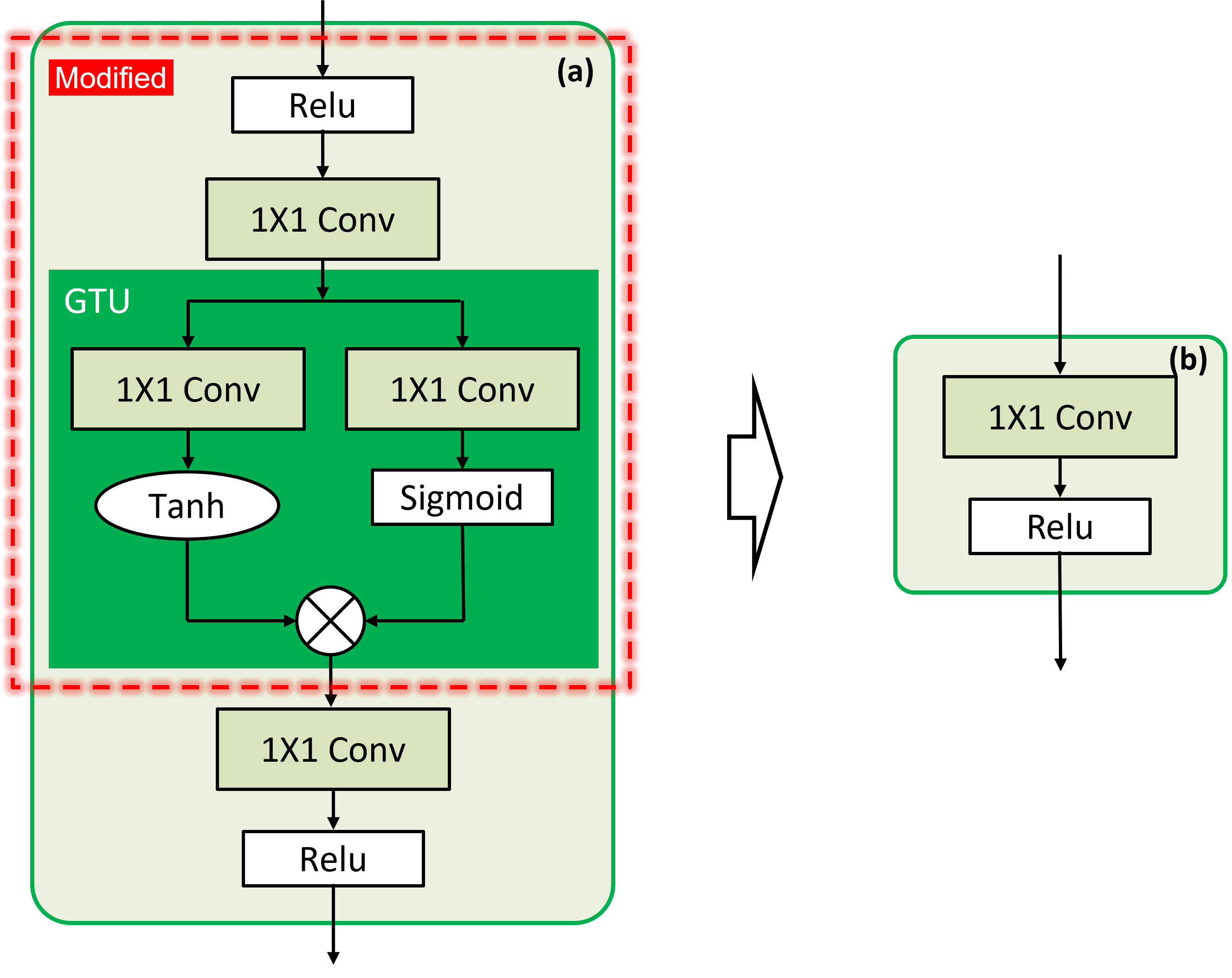}
\caption{(a) Original mask module and (b) modified mask module.}
\label{mask_module}
\end{figure}

Within the mask module, as depicted in Fig.~\ref{mask_module}, the Gate Tanh Unit (GTU) aids the model in selecting proper words or features for predicting subsequent words. However, its influence on the task at hand is relatively minimal. Consequently, to alleviate hardware overhead, we opt to eliminate this gating mechanism.

\begin{figure}[htbp]
\centering
\includegraphics[height=!,width=0.6\linewidth,keepaspectratio=true]{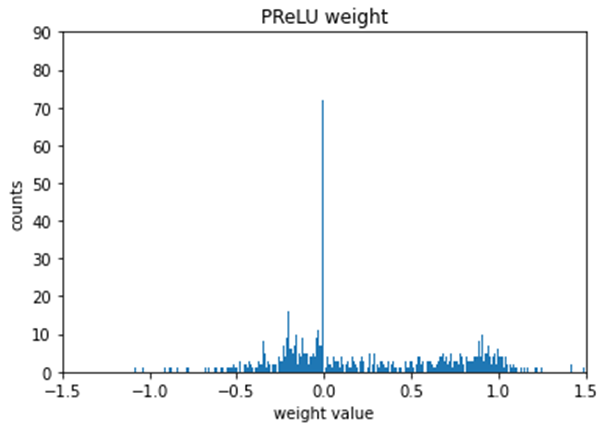}
\caption{The weight distribution of PReLU in this model.}
\label{prelu_weight}
\end{figure}

The model incorporates PReLU, enabling the neural network to adaptively learn nonlinearity across its layers. However, this introduces extra hardware demands. Fig.~\ref{prelu_weight} depicts the parametric weight distribution of PReLU for this model, revealing that a majority of the PReLU weights hover near zero. Consequently, values in the negative spectrum turn into weak tokens post PReLU processing. During inference, these weak tokens effectively become zero and have no significant influence on the final outcomes. To streamline the model, we substitute PReLU with ReLU.

\subsection{Streaming Aware Pruning}

\begin{figure}[htbp]
\centering
\includegraphics[height=!,width=1.0\linewidth,keepaspectratio=true]{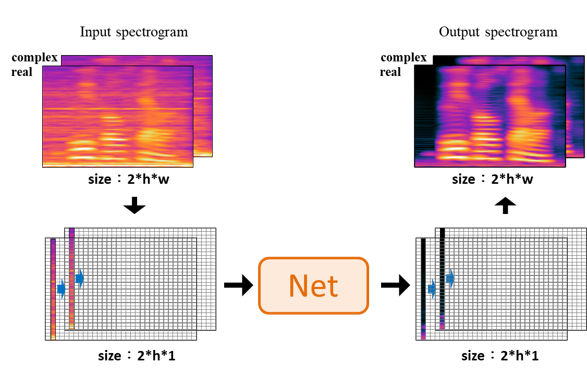}
\caption{The illustration of the streaming inference.}
\label{streaming_input}
\end{figure}

For effective streaming inference, it is essential that the model operates as a causal system, meaning it does not rely on future inputs. As illustrated in Fig.~\ref{streaming_input}, our model is designed to process a single time-step frame of the spectrogram in any given instance, catering to the streaming paradigm. Such an approach optimizes on-chip memory utilization, enabling the entirety of the model to execute on-chip, obviating the need for off-chip data access during intermediate stages. Moreover, leveraging this single time-step frame input (also termed as 1-D input) in conjunction with streaming prerequisites, we further refine the model to mitigate complexity. Specifically, this entails transitioning convolution kernels from 2-D to 1-D and incorporating subband attention, thereby eliminating the need for full-band multi-head attention.

In both the encoder and decoder, the Dilated Residual Block initially employs a 2-D convolution, driven by time-frequency input features. This setup hinders our ability to infer using a 1-D input suitable for streaming. Consequently, as depicted in Fig.~\ref{csp_res_block}, we modify the kernel size across all convolutions, transitioning from the 2-D kernel (2,3) to the streamlined 1-D kernel (1,5), a shift that aids in diminishing both computational complexity and memory requirements.

As illustrated in Fig.\ref{modify_trans_block} (a), a typical two-stage transformer block comprises two transformers: one dedicated to subband attention and the other to full-band attention. The approach, recognized as two-stage or dual-path signal processing, has gained traction in recent developments. Notably, the subband attention segment formulates attention mechanisms in sync with the spectrogram's frequency axis, based on the provided spectrogram input, posing no hindrance to streaming inference. Nonetheless, the full-band attention component, as seen in Fig.\ref{modify_trans_block} (a), formulates attention mechanisms aligned with the spectrogram's time axis. This poses a challenge for streaming inference, given that streaming input is restricted to a single time-step data.

Consequently, we eliminate the full-band multi-head attention module, which is highlighted in red in Fig.\ref{modify_trans_block} (a). In the case of the GRU within the full-band attention, we modify it from the bi-direction to the single direction, ensuring the module adheres to a causal system. The resultant modified transformer is depicted in Fig.\ref{modify_trans_block} (b).

\subsection{Hardware Friendly Model Design}
\label{section:Hardware friendly model design}
Building on the aforementioned structured pruning and alterations, the majority of blocks in TFTNN can be accelerated with parallel computing units. However,  both LN and softmax in the transformer block are not hardware-friendly, which requires online accumulations, leading to pronounced data dependencies. These dependencies hinder parallel execution, emerging as significant throughput bottlenecks. To build a hardware-friendly model, by leveraging the introduced cross-domain masking, we can resort to a constant BN for inference, bypassing the need for real-time LN computation. In addressing softmax, our proposition centers on the \textit{softmax-free MHA with an additional BN}, designed to eliminate data dependencies and facilitate an optimal matrix multiplication sequence. This is elaborated in the following.

\label{section:transformer with BN}

In TSTNN, both the encoder/decoder and the transformer employ LN. LN is regarded as a default normalization method in numerous speech tasks, primarily to compute the mean and variance of features in the channel dimension during inference. However, during model inference, LN requires real-time accumulation for both mean and variance, unlike the constant values used in BN. This means that the technique of folding BN into convolution is inapplicable here. Conversely, BN offers a faster inference than LN due to its fixed nature, allowing it to seamlessly fuse with convolution.

\begin{figure}[htbp]
\centering
\includegraphics[height=!,width=1.0\linewidth,keepaspectratio=true]{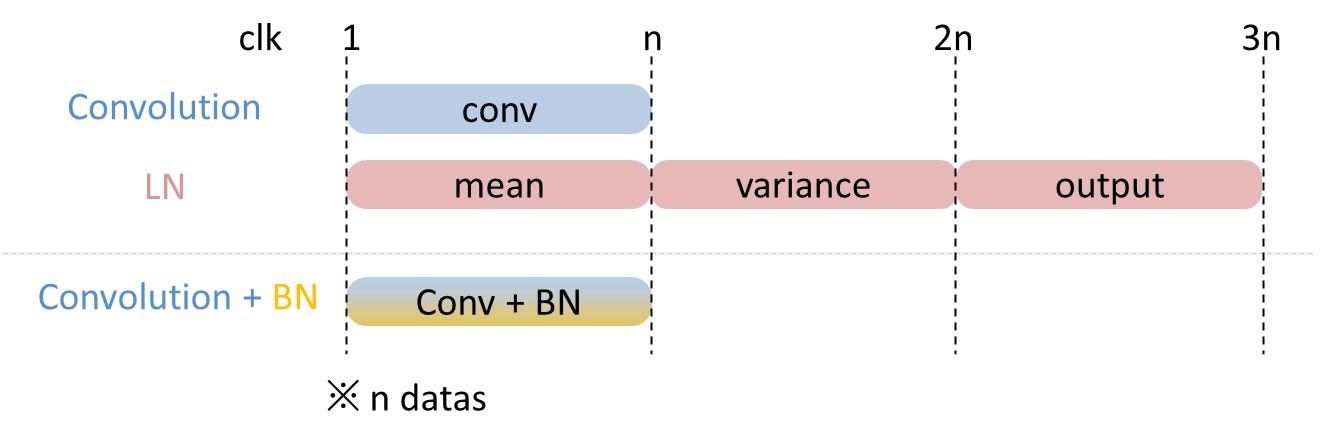}
\caption{Schedules for LN and BN.}
\label{layerNorm_flow}
\end{figure}

Given the time-frequency masking in this model, our input takes the form of a spectrogram, diverging from the traditional waveform. This spectrogram presents a reduced dynamic range compared to the original waveform and displays greater relevance across batches. Consequently, the adoption of BN, even in this altered context, doesn't lead to significant performance degradation. In light of its consistent value attribute and the minimal performance drop (as validated by simulation results), this study endorses the replacement of LN with BN throughout the model. This modifiction abolishes the need for online accumulations, slashing the LN cycle count by two-thirds, as delineated in Fig.~\ref{layerNorm_flow}.

\begin{figure}[htbp]
\centering
\includegraphics[height=!,width=1.0\linewidth,keepaspectratio=true]{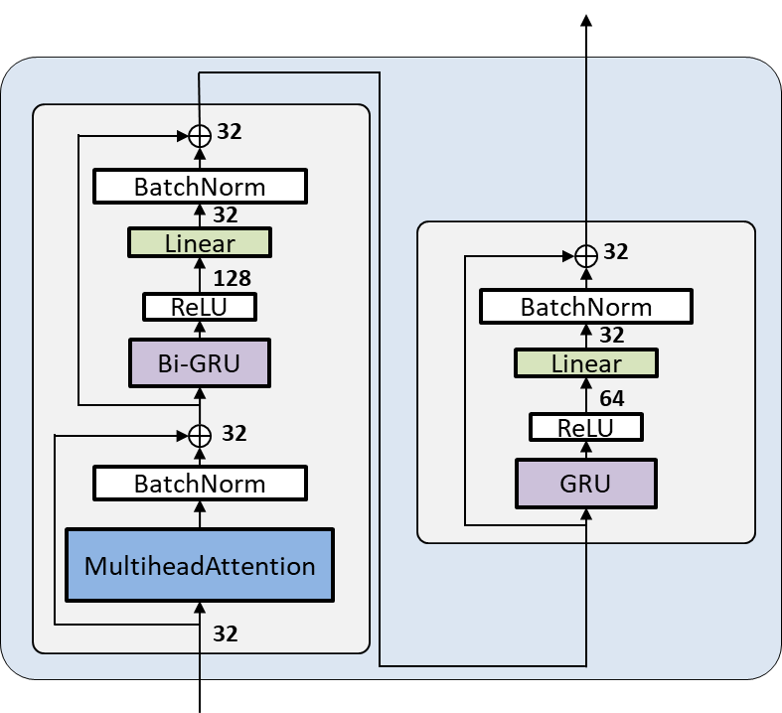}
\caption{The architecture of the proposed BN-based transformer block.}
\label{final_transformer_block}
\end{figure}

\begin{figure}[htbp]
\centering
\includegraphics[height=!,width=1.0\linewidth,keepaspectratio=true]{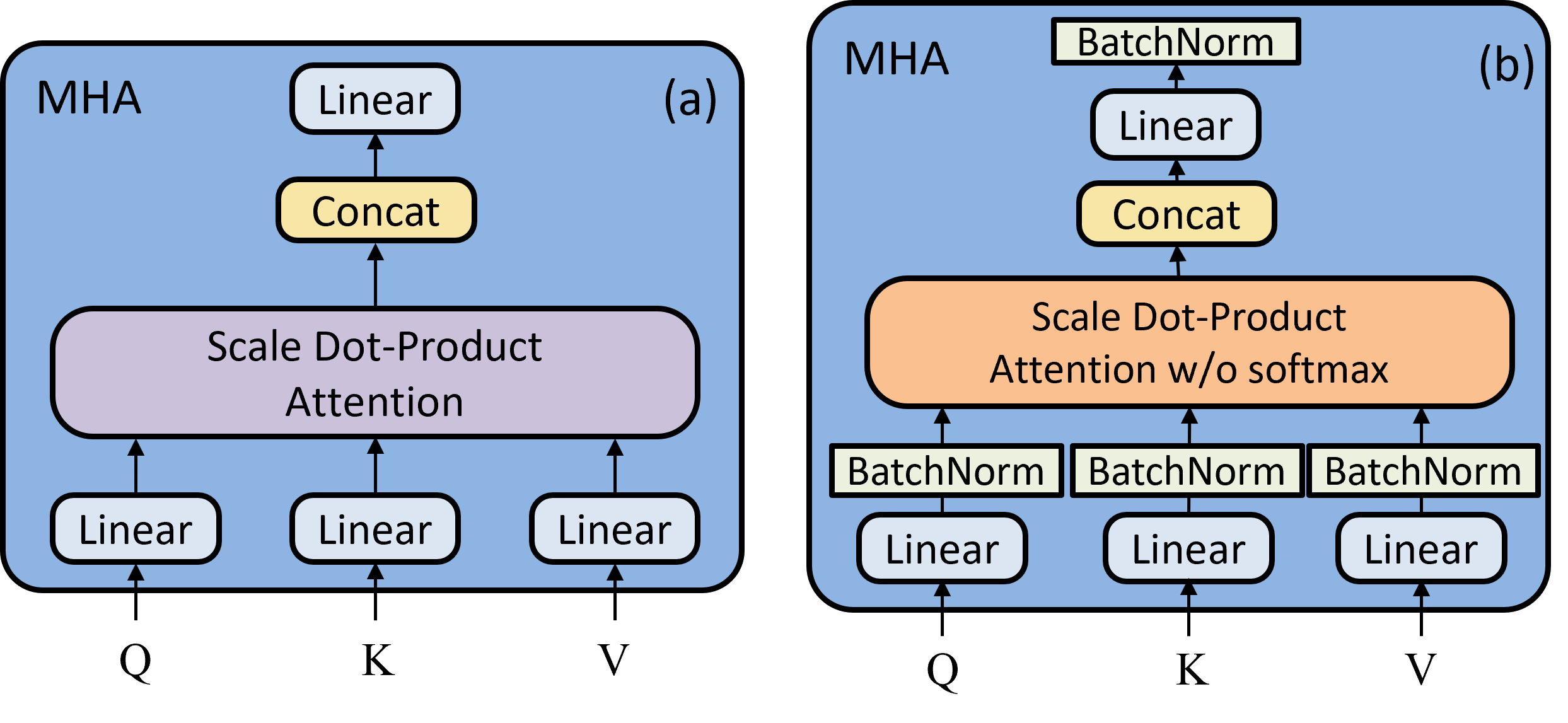}
\caption{(a) MHA and (b) softamx-free MHA with extra BN.}
\label{Softmax_free_MHA}
\end{figure}

However, a straightforward BN substitution within the transformer module can destabilize training. To counteract this, we've integrated an extra BN step into the MHA module, ensuring the normalization of FFN blocks and thwarting potential system crashes. The proposed BN-based transformer block, along with its MHA module, is illustrated in Fig.\ref{final_transformer_block} and Fig.\ref{Softmax_free_MHA}, respectively.

\begin{figure}[htbp]
\centering
\includegraphics[height=!,width=1.0\linewidth,keepaspectratio=true]{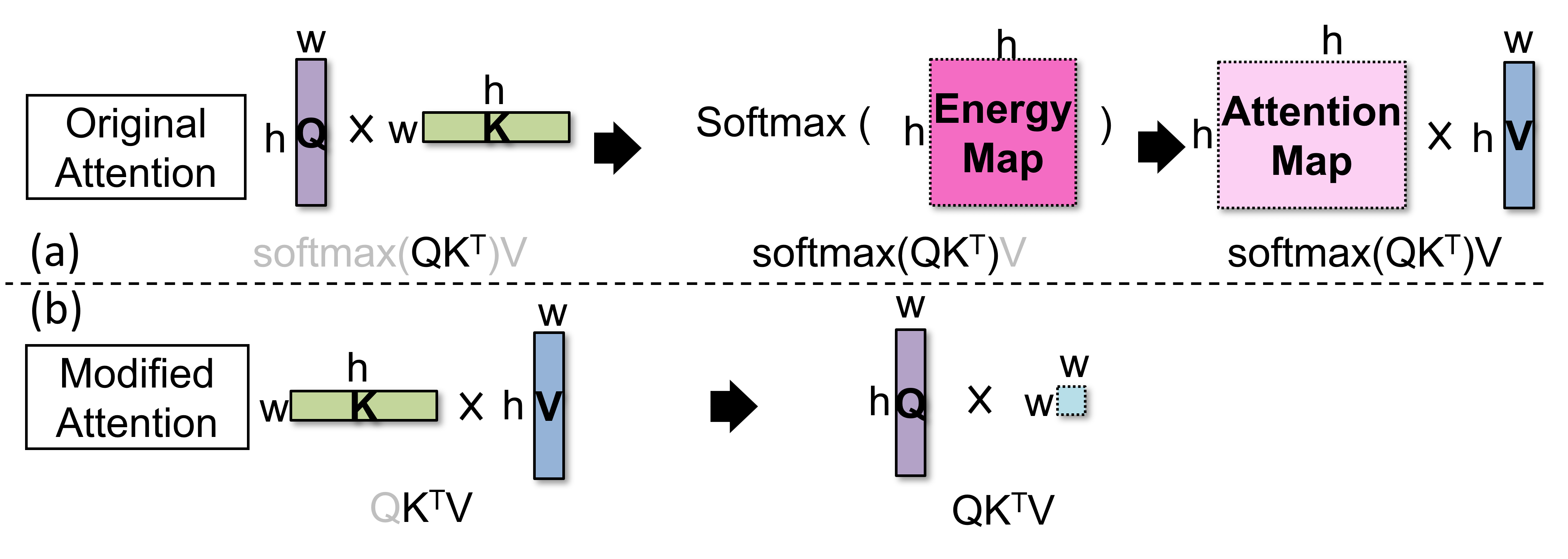}
\caption{Computation order of the MHA: (a)the original one, and (b) the proposed optimal order.}
\label{optimal_order_of_softmax}
\end{figure}

\begin{figure}[htbp]
\centering
\includegraphics[height=!,width=1.0\linewidth,keepaspectratio=true]{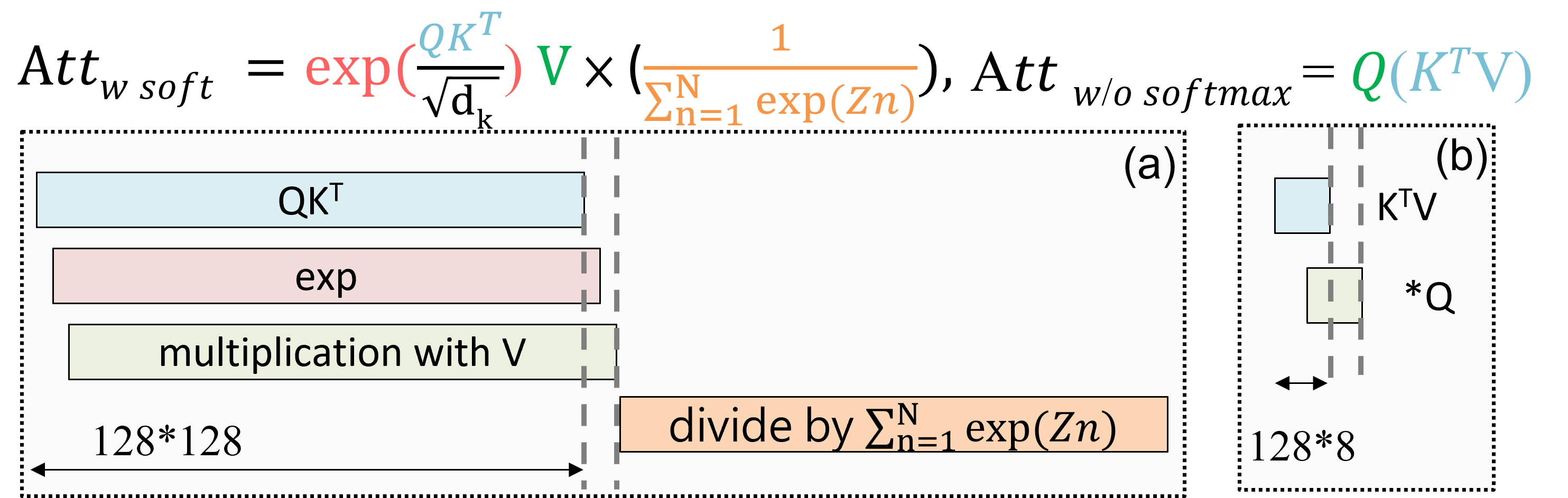}
\caption{The schedule of the attention (a) with and (b) without the softmax layer.}
\label{attention_schedule}
\end{figure}

As depicted in Fig.\ref{attention_schedule}, the MHA computation with softmax necessitates waiting and accumulating related values to derive the final results, leading to significant cycle consumption. Additionally, it demands significant memory allocation due to the storage requirements of the attention map, as illustrated in Fig.\ref{optimal_order_of_softmax} (a). Further compounding the issue, the inclusion of the exponential function in softmax poses an overhead for hardware design.

An earlier method presented in \cite{SimA} employs softmax-free attention, leveraging the L1 norm for Q and K, effectively removing softmax. However, the L1 norm isn't constant and mandates online computations during inference. Addressing this challenge, our proposal encompasses utilizing softmax-free attention, with Q and K normalized by BN, as demonstrated in Fig.\ref{Softmax_free_MHA} (b). This approach offers threefold advantages. Primarily, this BN is consistent with the LN replacement discussed earlier, incurring no added costs. Moreover, the BN values, being constant during inference, seamlessly integrate with linear convolutions. Finally, the omission of softmax facilitates the reformation of self-attention to attain the optimal matrix multiplication sequence, as depicted in Fig.\ref{optimal_order_of_softmax} (b). Within this structure, complexity is curtailed as outlined in Eq.\ref{eq:computation_ratio}, especially when the input length, $h=128$, is much larger than the input channel, $w=8$, in this model. Adopting this methodology, our hardware can enhance the attention operation speed by a factor of 16x (128/8), resulting in lower power consumption, as highlighted in Fig.\ref{attention_schedule}.

\begin{equation}
    \label{eq:computation_ratio}
    \frac{Complexity_{orig}}{Complexity_{new}}
    = \frac {(h \times w \times h) + (h \times h \times w)}
{(w \times h \times w) + (h \times w \times w)} =  \frac {h}{w}
\end{equation}

\subsection{The Final TFTNN Model}
\begin{figure}[htbp]
\centering
\includegraphics[height=!,width=1.0\linewidth,keepaspectratio=true]{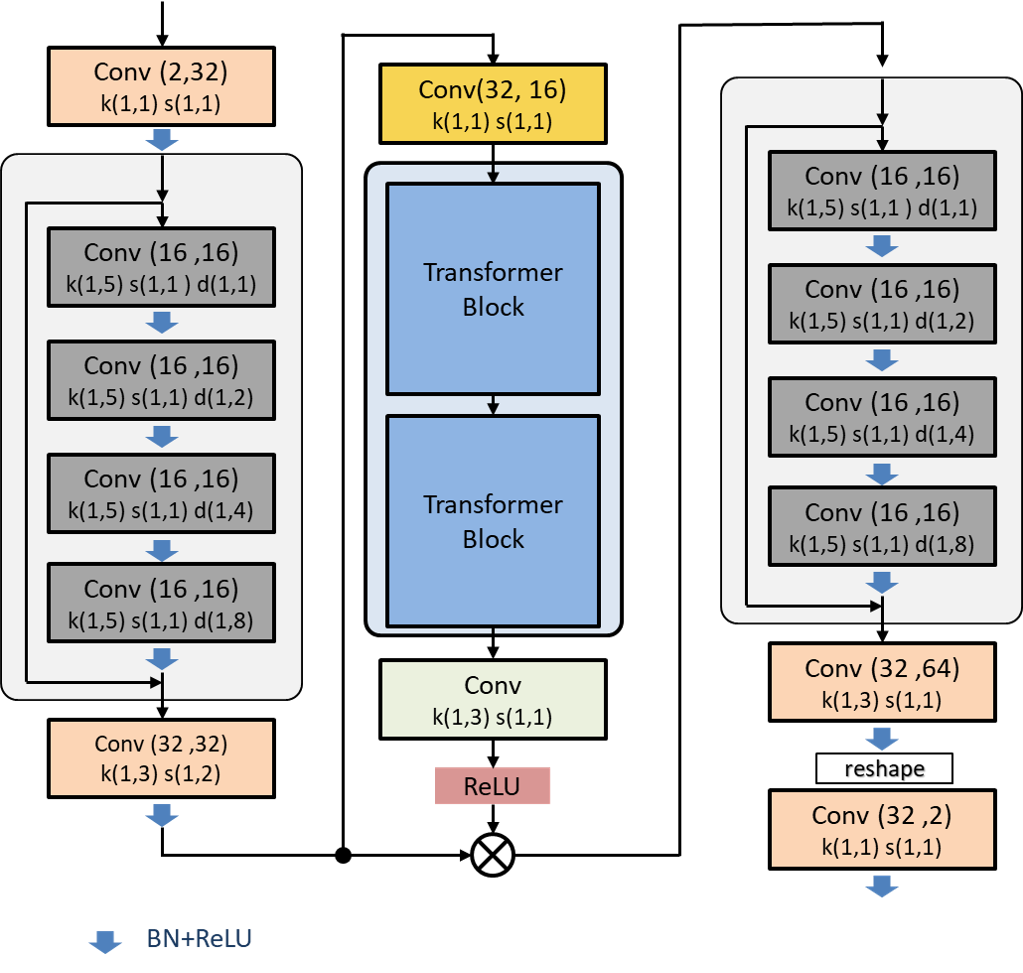}
\caption{The architecture of TFTNN with the transformer block in Fig.~\ref{final_transformer_block} and MHA in Fig.~\ref{Softmax_free_MHA} (b).}
\label{TSTNN_light}
\end{figure}

Building upon the aforementioned optimizations, we present our final TFTNN model in Fig.\ref{TSTNN_light}. This refined model consists of 55.92k parameters, marking an impressive reduction of approximately 94\% when compared to the original TSTNN.
In the transformer block illustrated in Fig.\ref{TSTNN_light}, we've modified the shortcut's location to facilitate a direct connection between BN and convolution, as detailed in Fig.\ref{final_transformer_block}. Additionally, given our utilization of BN for these layers, we've omitted the group normalization as evidenced in Fig.~\ref{modify_trans_block}.

\section{Hardware Accelerator}
\subsection{Design Analysis And Approach Overview}

Our design aims to process one frame within 16 ms, corresponding to 512 points for an 8K sampling rate with a 128-point hop length in STFT, thereby meeting the real-time streaming requirement. The computational demand for our lightweight model is set at 15.86 MMACs (million MACs) per frame. This can be addressed using 16 MACs, each running at a 1 MHz clock rate, resulting in an overall clock rate of 62.5 MHz for one-second data processing.

Despite the significant simplification in our proposed model, the hardware design still grapples with challenges like managing heterogeneous computing patterns, external memory access, and ensuring minimal power consumption. 

When it comes to heterogeneous computing patterns, our hardware must accommodate a range of computations, from convolution and GRU to multi-head attentions, a feature not commonly found in contemporary DLA designs. Most of these operations resemble either convolution or matrix multiplication processes. We address these challenges by breaking them down into element-wise MACs and arranging them in the proposed \textit{1-D processing array with adaptable SRAM addressing}, mitigating intricate hardware overheads. This 1-D array aligns seamlessly with our proposed 1-D streaming approach.

Additionally, the extensive DRAM power consumption from external memory access, attributable to deep layers and numerous shortcut connections, poses a challenge for low-power edge devices. However, our highly streamlined model permits the storage of all intermediate feature maps directly on-chip, eradicating the need for feature map I/O and facilitating the integration of shortcut connections. Consequently, only the primary input and final outputs necessitate feature map I/O.

Lastly, to further minimize power consumption, we incorporate a gating mechanism into computational units and buffers, extending our efforts beyond the aforementioned complexity reduction. Further details are elaborated upon in the subsequent sections.

\subsection{System Architecture}
\begin{figure}[htbp]
\centering
\includegraphics[height=!,width=1.0\linewidth,keepaspectratio=true]{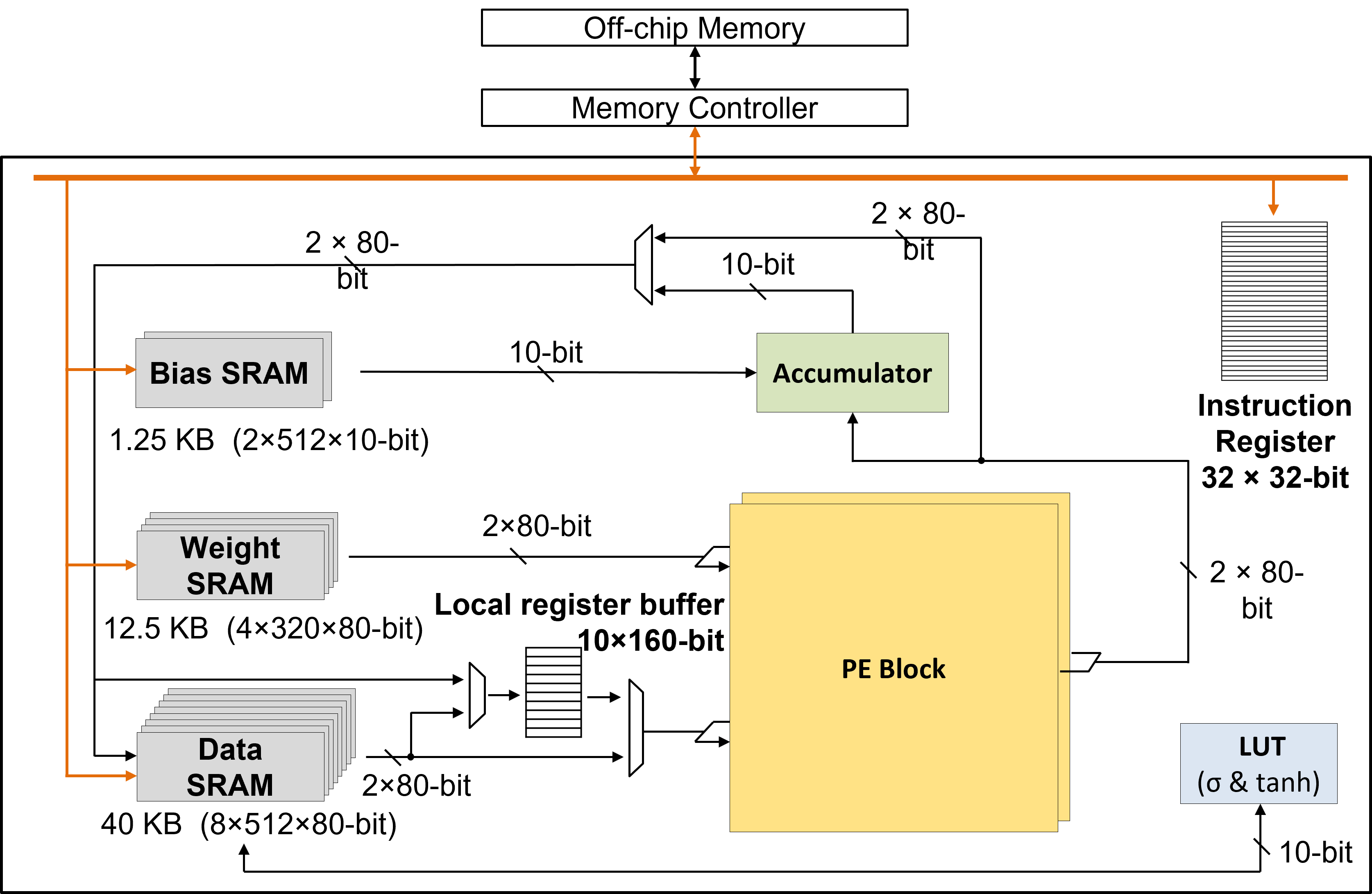}
\caption{The system architecture.}
\label{system_architecture}
\end{figure}

\subsubsection{Overview}
The proposed system architecture for our model is depicted in Fig.~\ref{system_architecture}. Our hardware processes the model sequentially, layer by layer. To curtail external memory access, we store all intermediate feature maps directly in the on-chip SRAM. This strategy means that only the initial input and the final outputs require interaction with off-chip memory. However, the constraints of on-chip buffers necessitate weight adjustments during inference. For seamless operation, all SRAMs in our setup are designed as ping-pong buffers. The system interfaces with external memory via a memory controller, equipped with distinct 80-bit data input and output channels. The entire system can be programmed using custom instructions.

\subsubsection{On-chip buffer}
Buffers dedicated to data, weight, and bias are allocated into 8, 4, and 2 banks respectively, aligning with the bandwidth demands of a single PE block. The SRAM data's word line is aligned with the signal's length, and the word line for the weight SRAM is tailored to fit the kernel size and the output channel. This arrangement facilitates the sequential access of the SRAM during convolution operations.

To minimize SRAM access and conserve power, our system integrates 10 local register buffers, facilitating data exchange between SRAM and PE blocks, as illustrated in Fig.~\ref{system_architecture}. Each local register buffer encompasses 160 bits, aligning with the bandwidth capacity of a pair of PE blocks. Through these buffers, data reuse is streamlined, eliminating frequent SRAM access during operations. Furthermore, these buffers are adept at storing intermediary outcomes within the GRU process.

\subsubsection{PE block}
\label{section:PE_block}

\begin{figure}[htbp]
\centering
\includegraphics[height=!,width=1.0\linewidth,keepaspectratio=true]{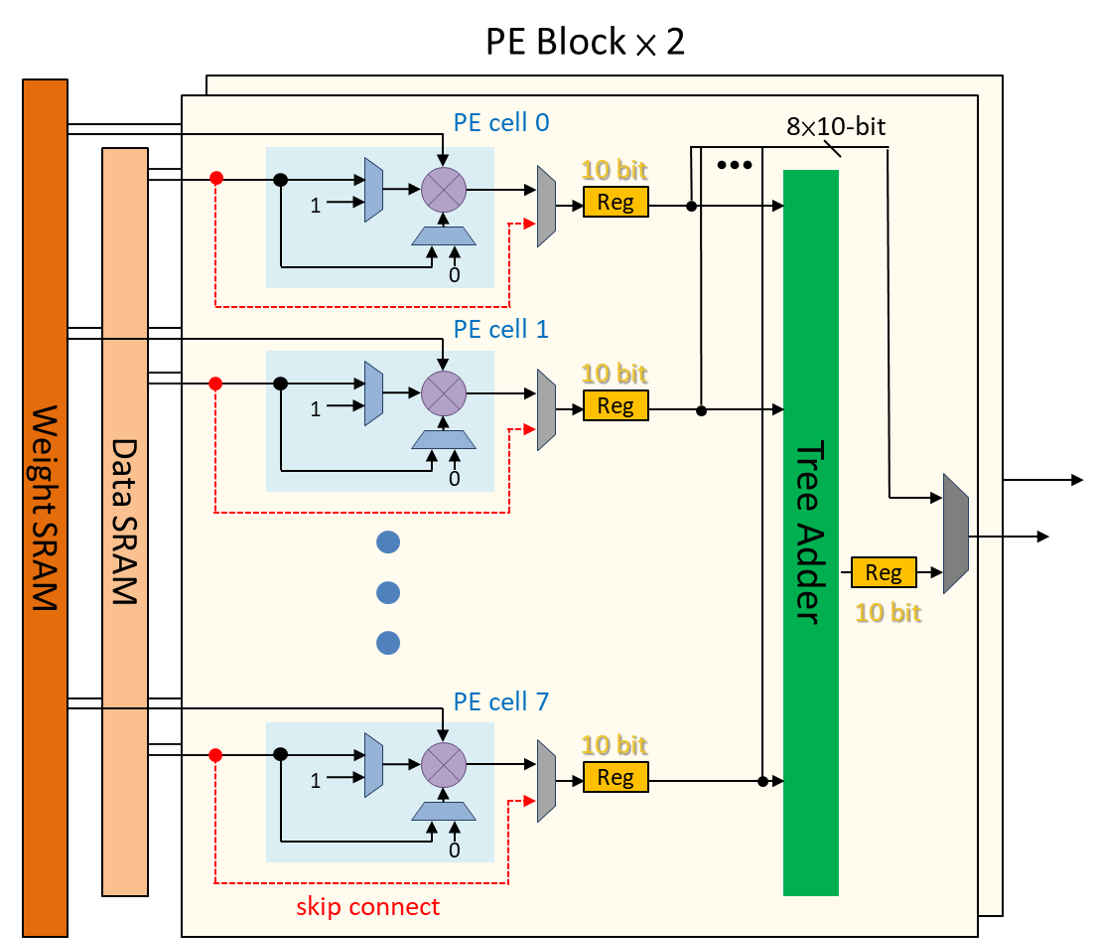}
\caption{The PE block architecture.}
\label{PE}
\end{figure}

For our streamlined model, we utilize two PE blocks, each equipped with eight element-wise MACs, for the eight input channels. Our design can operate up to 16 MACs simultaneously. The results from the PE blocks are aggregated, with an accompanying bias integrated during this accumulation. The architecture of the PE block, consisting of 8 PE cells and a tree adder, is illustrated in Fig.~\ref{PE}. If the input data during convolution is zero, the PE computation is skipped, and data is sent directly to the tree adder, optimizing energy consumption. The PE block is versatile and can be reconfigured to accommodate all operations mandated by our model, including convolutions, transformer mechanisms, shortcut additions, and masking. In convolution operations, the PE cell is set to multiplication mode, and the tree adder consolidates all outputs from the PE cells. The tree adder's output feeds into the accumulator for further summation. In addition, the PE block is equipped to manage the shortcut and mask layer through element-wise additions and multiplications. This method of element-wise multiplication and addition also applies when calculating the dot product in the attention layer.

\subsection{Data Flow With Configurable SRAM Addressing}
Our model's primary operations are broadly classified into two types: convolution and matrix multiplication. The convolution flow is applied in the convolution layer of the encoder and decoder, and the linear layer of GRU and MHA. The attention layer of MHA employs the matrix multiplication flow. The gate operations of GRU, involving element-wise multiplications and additions, follow a flow akin to matrix multiplication. Both of these operations are effectively handled using the proposed 1-D array. Subsequently, we will detail the data flow for convolution and matrix multiplication, and then describe how these are integrated into GRU and MHA in a systematic manner. To streamline hardware design, different data flows can be effectively managed with straightforward control in conjunction with varying SRAM addressing.
\subsubsection{Channel-wise input flow for convolution}
\begin{figure}[htbp]
\centering
\includegraphics[height=!,width=1.0\linewidth,keepaspectratio=true]{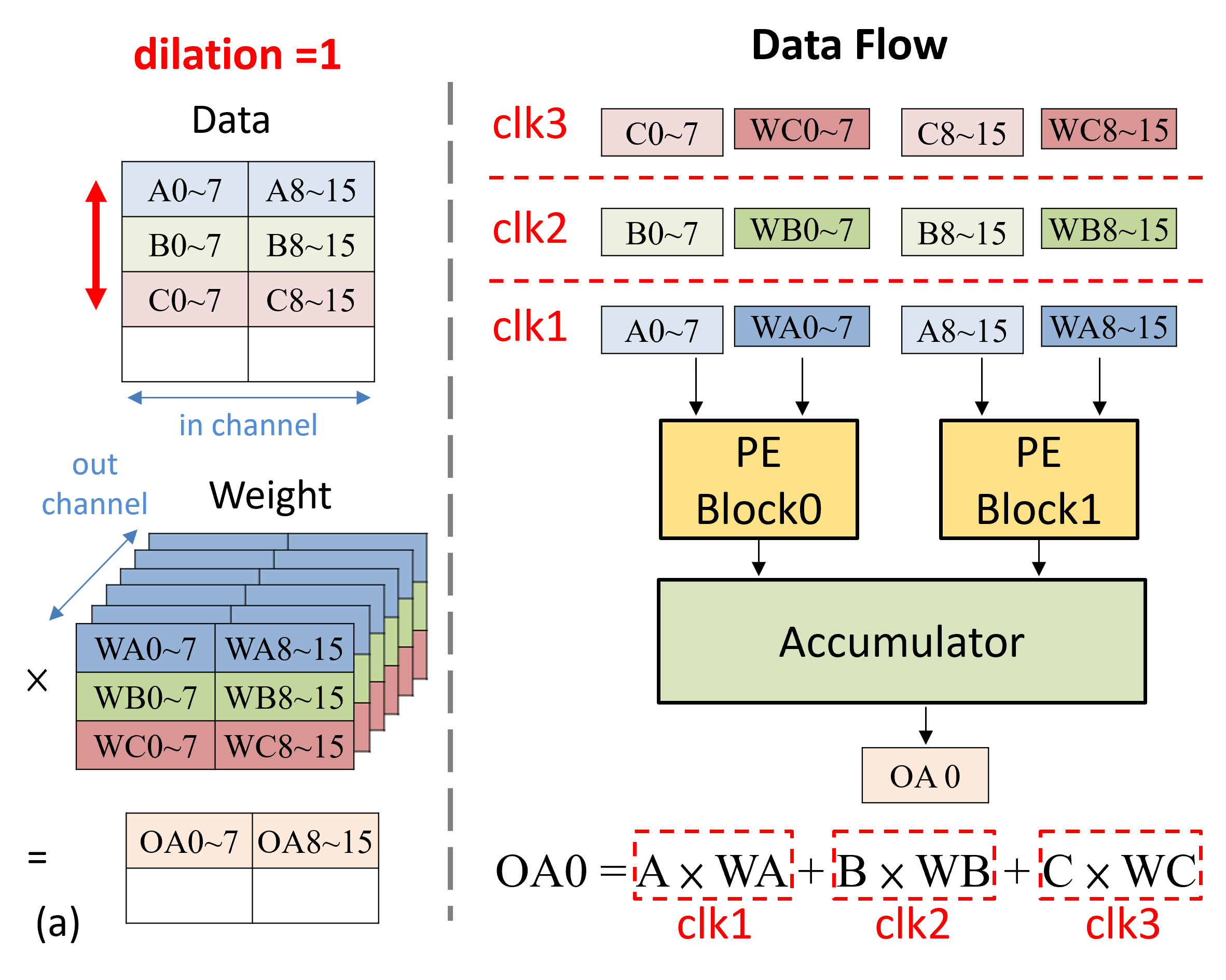}
\includegraphics[height=!,width=1.0\linewidth,keepaspectratio=true]{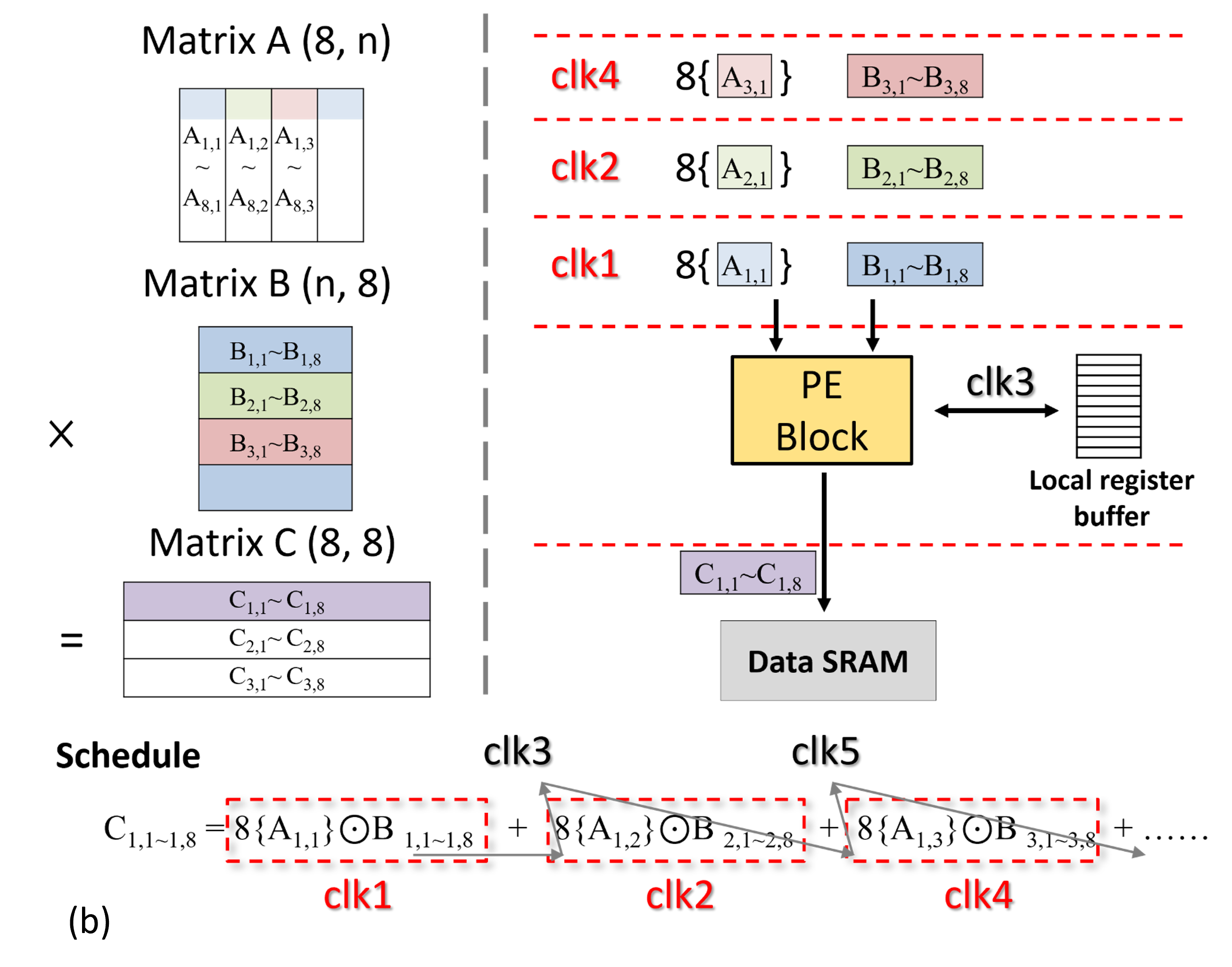}
\caption{The data flow of the (a) convolution, and (b) matrix multiplication operation.}
\label{convolution_flow}
\end{figure}

For a unified data flow that performs a 1-D convolution across various kernel sizes and dilation, we employ a channel-wise input approach, illustrated in Fig.~\ref{convolution_flow}(a). When the dilation of the convolution surpasses 1, it necessitates accessing different positions within the data SRAM. In this data flow approach, input data sharing the same kernel dimension are relayed to the PE blocks each cycle. The PE blocks then calculate the product and consolidate results along the channel dimension. These interim sums are then compiled by the accumulator to produce the final output.

\subsubsection{Data flow for matrix multiplication}
\label{section:matrix_multiplication}

Contrary to CNNs, the MHA approach processes matrix multiplication along the signal length dimension. However, signals are sequentially stored in SRAM, hindering parallel access. One solution might involve altering the SRAM data layout, but this complicates control and adds overhead.

To address this issue, we employ the element-wise MAC for matrix multiplication, as depicted in Fig.~\ref{convolution_flow}(b). Using two matrices of array size 8$\times$n as an illustration—where 8 represents the \textit{embedding} dimension of MHA and \textit{n} denotes the signal size—both Matrix A and Matrix B reside in separate banks of the data SRAM. Data from $A_{1,1}$ and $B_{1,1-1,8}$ is accessed, with $A_{1,1}$ broadcasted to all PE cells in the PE blocks. This facilitates the element-wise multiplication with $B_{1,1-1,8}$. This computation process is consistently applied to subsequent data. Moreover, data accumulation is performed within the same PE block, with the partial sum stored in the local register buffer.

\subsubsection{GRU operation}

\begin{figure}[htbp]
\centering
\includegraphics[height=!,width=1.0\linewidth,keepaspectratio=true]{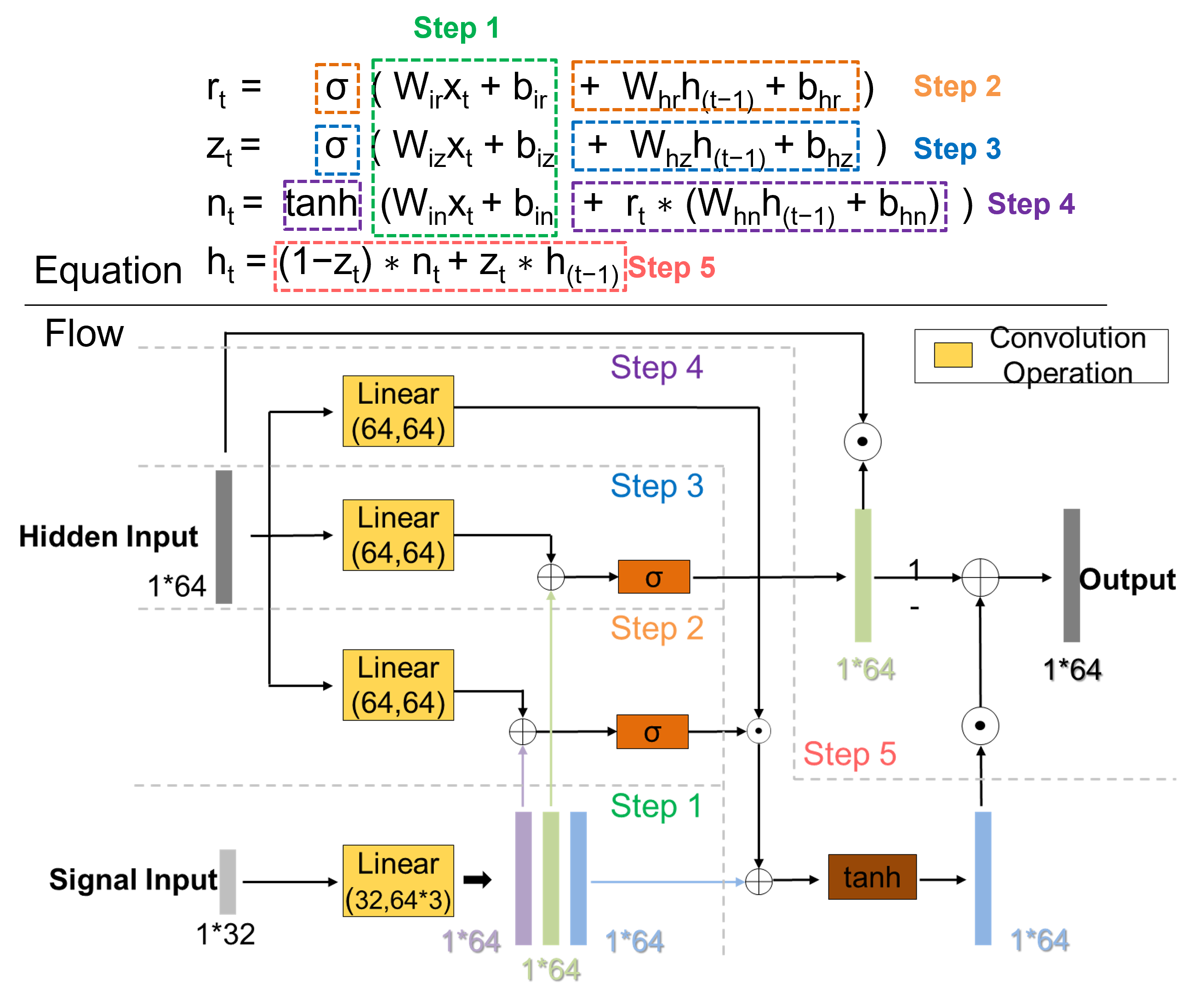}
\caption{The operation steps of the GRU.}
\label{gru_data_flow}
\end{figure}

As depicted in Fig.~\ref{gru_data_flow}, there are five sequential GRU operation steps designed for hardware implementation. Each of these steps applies either the convolution or matrix multiplication flow, depending on its specific requirements. The initial step involves computing the three linear layers associated with the input layer. Subsequently, the three following steps calculate the reset gate, update gate, and new gate, employing element-wise multiplication and addition akin to the matrix multiplication process. Additionally, a Lookup Table (LUT) is utilized to manage the sigmoid and tanh functions. The concluding step yields the new hidden state result.

Given that our system is equipped with only two PE blocks, these stages function in a sequential manner rather than in parallel. The PE blocks sequentially traverse all five steps to process a single data input, then iterate over the entire signal.

\subsubsection{MHA operation}
\begin{figure}[htbp]
\centering
\includegraphics[height=!,width=1.0\linewidth,keepaspectratio=true]{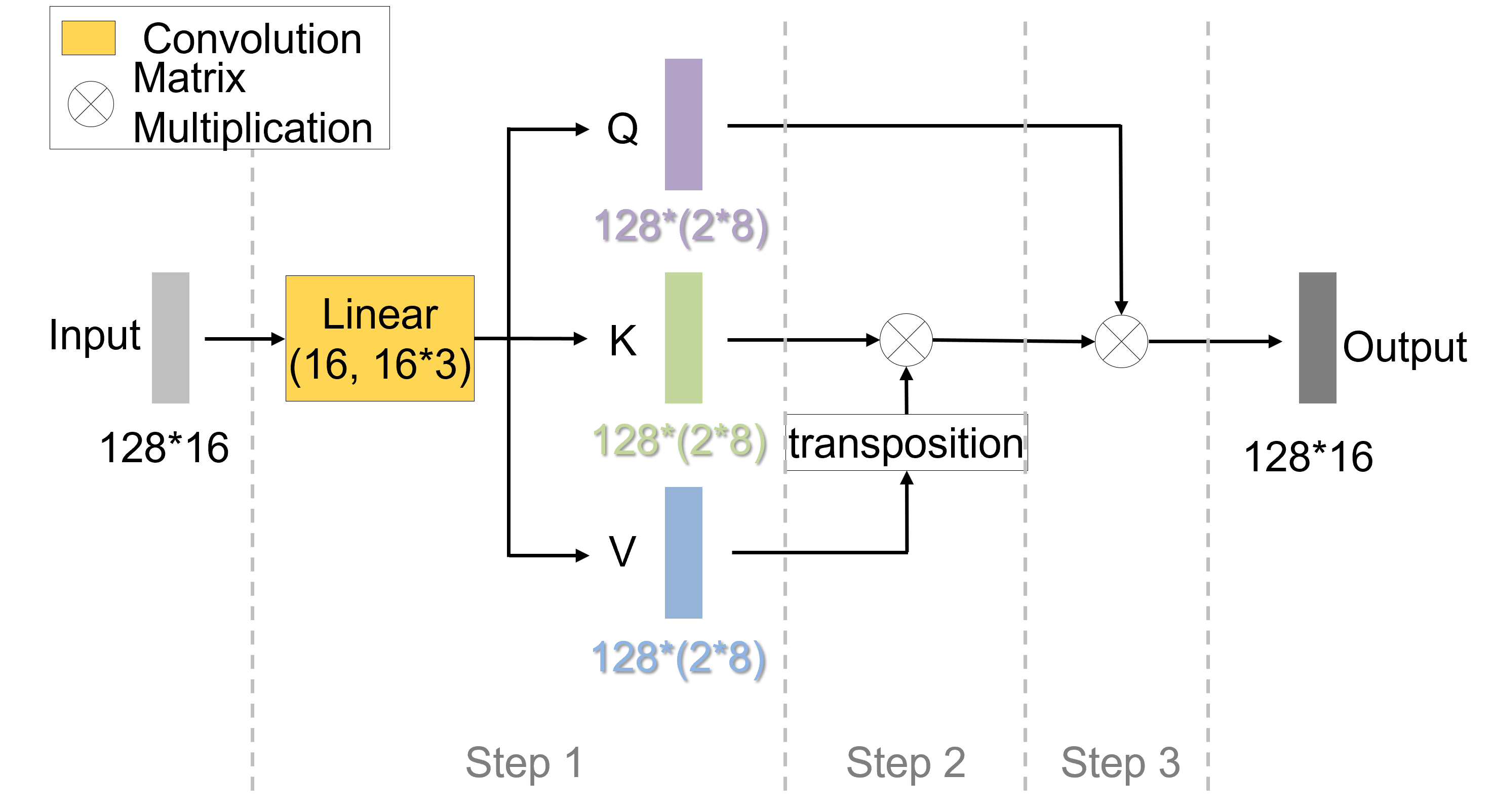}
\caption{The operation steps of the MHA.}
\label{mha_data_flow}
\end{figure}

As depicted in Fig.~\ref{mha_data_flow}, there are three designated MHA operation steps tailored for hardware execution. Each of these steps utilizes either the convolution or matrix multiplication flow, depending on the specific requirement. Consider an input signal characterized by a length of 128 and an \textit{embedding} dimension of 16. In the initial step, the linear layer in self-attention is computed for query(Q), key(K), and value(V). In the subsequent stage, only the matrix multiplication of key(K) and value(V) is computed, due to the omission of the softmax module. The third step involves the dot product computation with query(Q). This progression delineates the most efficient computation sequence for MHA in our design. The PE blocks sequentially process these steps for each input.

\section{Experimental Results}
\begin{table*}[htbp]
\begin{center}
\caption{Performance comparison with the state-of-the-art works in recent years.}
\label{table:comparison with state of art}

\begin{tabular}{ccccccccccc}
\hline
\multirow{2}{*}{\textbf{}} &
  \multirow{2}{*}{\textbf{Dataset}} &
  \textbf{clean} &
  \multicolumn{6}{c}{\textbf{VoiceBank}} &
  \multirow{3}{*}{\textbf{Parameters}} &
  \multirow{3}{*}{\begin{tabular}[c]{@{}c@{}}\textbf{Computations}\\ \textbf{(GMac)}\end{tabular}} \\ \cline{3-9}
     &            & \textbf{noise} & \multicolumn{3}{c}{\textbf{UrbanSound8K}}         & \multicolumn{3}{c}{\textbf{DEMAND}}   &         &       \\ \cline{1-9}
Year & Model      & Domain         & PESQ           & STOI           & SNR             & PESQ           & STOI           & SNR    &         &       \\ \hline
2019 & ConvTasNet\cite{ConvTasNet_2019} & T  & 2.051 & 0.803 & 11.470 & 2.945  & 0.889 & 15.564 & 3.986 M & 8.34  \\
2020 & DCCRN\cite{DCCRN}      & T  & 2.556 & 0.857 & 14.491 & 3.302  & 0.923 & 17.527 & 2.495 M & 3.26  \\
2021 & TSTNN\cite{TSTNN_2021}      & T  & 2.637 & 0.869 & 14.622 & 3.452  & 0.939 & 17.189 & 922.9 k & 9.87  \\
2022 & DBT\_Net\cite{DBT_Net}   & TF & \textbf{2.909} & \textbf{0.891} & \textbf{15.228} & \textbf{3.658} & \textbf{0.951} & 17.922 & 2.908 M & 20.08 \\ \hline
&
TFTNN 
&TF &2.746 &0.878 &14.745 &3.501 &0.939 &\textbf{18.545} &\textbf{55.9 k} &\textbf{0.496} \\ \hline
\end{tabular}

\end{center}
\end{table*}
\begin{table*}[htbp]
\centering
\renewcommand{\arraystretch}{}
\caption{Ablation analysis of different domain methods.}
\label{table:comparison of the different domain methods}
\begin{tabular}{ccccccccc}
\hline
\multirow{2}{*}{\textbf{Dataset}}      & 
\multicolumn{4}{c}{\textbf{clean}}     & 
\multicolumn{3}{c}{\textbf{VoiceBank}} & 
\multirow{3}{*}{\textbf{\begin{tabular}[c]{@{}c@{}}Parameter\\ (k)\end{tabular}}} \\ & 
\multicolumn{4}{c}{\textbf{noise}} & 
\multicolumn{3}{c}{\textbf{UrbanSound8K}} &  \\ \cline{1-8}

Model   & 
\begin{tabular}[c]{@{}c@{}}Mask \\ Domain\end{tabular} & Loss & FFT  & \begin{tabular}[c]{@{}c@{}}hop\\ length\end{tabular} & PESQ   & STOI    & SNR    &    \\ \hline

\multirow{4}{*}{TSTNN}  
&  T     & T+F      & -     & -     & 2.6390  & 0.8680  & 14.9370   & \multirow{4}{*}{922.9}\\
& TF    & F        & 400    & 160   & 2.6788  & 0.8650  & 14.9650   &  \\

& \multirow{2}{*}{TF}   & \multirow{2}{*}{T+F} & 400   & 160   & 2.7970  & 0.8740 & 14.6160  &    \\

& &      & 512   & 128   & \textbf{2.8848} & \textbf{0.8839} & \textbf{15.0324} &  \\ \hline

\multirow{2}{*}{TFTNN}   & TF    & F     & \multirow{2}{*}{512} & \multirow{2}{*}{128}                                 & 2.1190          & 0.8008          & 11.0864          & \multirow{2}{*}{55.9}   \\
& TF   & T+F      &    &    & 2.7460          & 0.8780          & 14.7450          &     \\ \hline
\end{tabular}

\end{table*}

\begin{table}[htbp]
\centering
\renewcommand{\arraystretch}{}
\caption{Ablation analysis of the use of different transformer block numbers.}
\label{table:comparison of the different transformer numbers}
\begin{tabular}{ccccc}
\hline
\multirow{2}{*}{\textbf{Dataset}} & \textbf{clean}  & \multicolumn{3}{c}{\textbf{VoiceBank}}   \\ & \textbf{noise}  & \multicolumn{3}{c}{\textbf{UrbanSound8K}}  \\ \hline
Model  & trans block num & PESQ & STOI & SNR  \\ \hline
\multirow{4}{*}{TFTNN}     & 4               & \textbf{2.7698} & \textbf{0.8801} & 14.5938          \\
                                  & 3               & 2.7263          & 0.8743          & 13.7332          \\
                                  & 2               & 2.7459          & 0.8779          & \textbf{14.7450} \\
                                  & 1               & 2.5974          & 0.8602          & 14.0212          \\ \hline
\end{tabular}

\end{table}

\begin{table}[htbp]
\caption{Ablation analysis of the transformer with LN and extra BN in MHA for TSTNN.}
\label{table:comparison of Transfomer with Batch Normalization and Layer Normalization}
\begin{tabular}{cccccccc}
\hline
\multirow{2}{*}{Dataset}  & \textbf{clean} & \multicolumn{6}{c}{VoiceBank}                                                                                                                                                                               \\ \cline{2-8} 
                          & \textbf{noise} & \multicolumn{3}{c}{UrbanSound8K}                                                                     & \multicolumn{3}{c}{DEMAND}                                                                           \\ \hline
Ex. BN                  & Norm                            & PESQ                            & STOI                            & SNR                              & PESQ                            & STOI                            & SNR                              \\ \hline
                          & LN                              & \textbf{2.797} & 0.874                           & \textbf{14.616} & 3.491                           & 0.939                           & \textbf{18.936} \\ \cline{2-2}
                          & \multirow{2}{*}{BN}             & 2.724                           & 0.872                           & 14.468                           & 3.465                           & 0.941                           & 18.488                           \\ \cline{1-1}
\checkmark &                                 & 2.792                           & \textbf{0.876} & 14.301                           & \textbf{3.526} & \textbf{0.943} & 18.168                           \\ \hline
\end{tabular}
\end{table}

\begin{table*}
\centering
\caption{Comparisons with other designs.}
\label{table:hardware_result}
\begin{tabular}{|c|c|c|c|c|c|} 
\hline
\rowcolor[rgb]{0.753,0.753,0.753}                                                                 & \textbf{This work}                                              & \cite{ISSCC_2017_asr}                                                  & \cite{ISSCC2021_denoising}                                                                         & \cite{JSSC_2020}                                                         & \cite{NTUADSP}               \\ 
\hline
Application                                                                                       & \begin{tabular}[c]{@{}c@{}}Speech \\Enhancement\end{tabular}    & \begin{tabular}[c]{@{}c@{}}Speech  \\Recognition\end{tabular} & \begin{tabular}[c]{@{}c@{}}Speech \\Recognition\end{tabular}                         & \begin{tabular}[c]{@{}c@{}}Speech \\Recognition\end{tabular}        & Hearing devices                       \\ 
\hline
Algorithm                                                                                         & \begin{tabular}[c]{@{}c@{}}CNN, GRU, \\Transformer\end{tabular} & CNN                                                           & \begin{tabular}[c]{@{}c@{}}Attention, \\based RNN\end{tabular}                       & LSTM, RNN                                                           & CNN-FC                 \\ 
\hline
SRAM(KB)                                                                                          & 53.75                                                           & 730                                                           & 10035                                                                                & 297                                                                 & 327                         \\ 
\hline
Technology (nm)                                                                                   & 40                                                              & 65                                                            & 16                                                                                   & 65                                                                  & 40                        \\ 
\hline
\textsuperscript{e}Area(KGE) (logic only)                                                         & 207.8                                                           & 2088                                                          & -                                                                                    & -                                                                   & 6200                         \\ 
\hline
Area(mm\textsuperscript{2})                                                                       & \textsuperscript{c}0.367                                        & \textsuperscript{c}9.61                                       & \textsuperscript{c}8.84                                                              & \textsuperscript{d}7.74                                             & 4.2  \\ 
\hline
Supply Voltage (V)                                                                                & 0.9                                                             & 0.6-1.2                                                       & 0.55-1.0                                                                             & 1.1                                                                 & 0.6                       \\ 
\hline
Frequency (MHz)                                                                                   & 62.5 – 250                                                      & 3-86                                                          & 130-775                                                                              & 8-80                                                                & 5                      \\ 
\hline
Precision                                                                                         & FP10                                                            & -                                                             & FP8                                                                                  & \begin{tabular}[c]{@{}c@{}}FP6 (weight)  \\FP13 (Act.)\end{tabular} & INT16                      \\ 
\hline
Power (mW)                                                                                        & 8.08 - 20.1                                                     & 1.8-7.8                                                       & 19 – 227                                                                             & 67.3                                                                & 2.17                    \\ 
\hline
PE number                                                                                         & 16                                                              & 32                                                            & 1024                                                                                 & 65                                                                  & 64                       \\ 
\hline
Throughput (GOPS)                                                                                 & 2 - 8                                                           & 0.019-2.7                                                     & 148.2-590.2                                                                          & 24.6                                                                & -                       \\ 
\hline

\multirow{2}{*}{Energy eff. (TOPS/W)} & \multirow{2}{*}{0.248 - 0.398} & 0.01-0.34 & 2.6-7.8 & 2.45 & 1.2 \\ 
\cline{3-6}
 &  & \textsuperscript{a}0.03-0.24 & \textsuperscript{a}1.16-1.28 & \textsuperscript{a}5.95 & \textsuperscript{a}0.53 \\ 
\hline
\multirow{2}{*}{\begin{tabular}[c]{@{}c@{}}Area eff. \\(GOPS/mm\textsuperscript{2})\end{tabular}} & \multirow{2}{*}{5.45 - 21.798~} & 0.001-0.28 & 16.76-66.76 & 3.23 & \multirow{2}{*}{0.62~} \\ 
\cline{3-5}
 &  & \textsuperscript{b}0.003-0.202 & \textsuperscript{b}8.29-9.97 & \textsuperscript{b}5.24 &  \\ 
\hline

\multicolumn{4}{|l}{\begin{tabular}[c]{@{}l@{}}\textsuperscript{a}Normalized energy efficiency = energy efficiency × ( process/40nm ) × ( voltage/0.9V )\textsuperscript{2}. \\\textsuperscript{c}Core only size. \\\textsuperscript{e}The area is shown in terms of the size of the kilo NAND2 gates (KGE).\end{tabular}}   & \multicolumn{2}{l|}{\begin{tabular}[c]{@{}l@{}}\textsuperscript{b}Technology scaling ( process / 40nm ) \\\textsuperscript{d}Chip size. \\\end{tabular}}  \\
\hline
\end{tabular}
\end{table*}

\subsection{Experimental setup}
In our experiments, we employed the widely-used VoiceBank+DEMAND dataset to evaluate the performance of TFTNN. The clean signals sourced from this dataset originate from the Voice Bank corpus\cite{VoiceBank}. This corpus has more than 300 hours of recordings and comprises 11,572 utterances spoken by 28 unique speakers, with an equal distribution of 14 males and 14 females, designated for training. The test set encompasses 824 utterances, delivered by 2 speakers, one male and one female.

To introduce diversity and challenge, we incorporated a secondary noise dataset - UrbanSound8k \cite{UrbanSound8K}. The UrbanSound8k dataset is a rich collection of 8,732 urban sound snippets distributed across 10 distinct categories. By mixing the Voice Bank Corpus with UrbanSound8k at a Signal-to-Noise Ratio (SNR) of 2.5 dB, we ensured that our model is exposed to a varied and intricate noisy speech environment during evaluations.

We utilize three evaluation metrics to assess our model: Perceptual Evaluation of Speech Quality (PESQ)~\cite{pesq}, Short-Time Objective Intelligibility (STOI)~\cite{stoi}, and Signal-to-Noise Ratio (SNR)~\cite{SNR}. The score ranges for these metrics are as follows: -0.5 to 4.5 for PESQ, 0 to 1 for STOI, and -10 to 35 for SNR. For all these metrics, a higher score signifies superior speech quality.

During data preprocessing, we employ an 8K sampling rate for all utterances. For the STFT, the length is set to 512 and the hop length to 128, which translates to durations of 64ms and 16ms, respectively. To mitigate signal edge disparities and reduce Fourier transform leakage, we utilize the Hanning window. For both training and testing phases, speech segments of 3 seconds are considered. If an utterance exceeds 3 seconds, a random 3-second segment from the utterance is chosen.

\begin{figure}[htbp]
\centering
\includegraphics[height=!,width=1.0\linewidth,keepaspectratio=true]{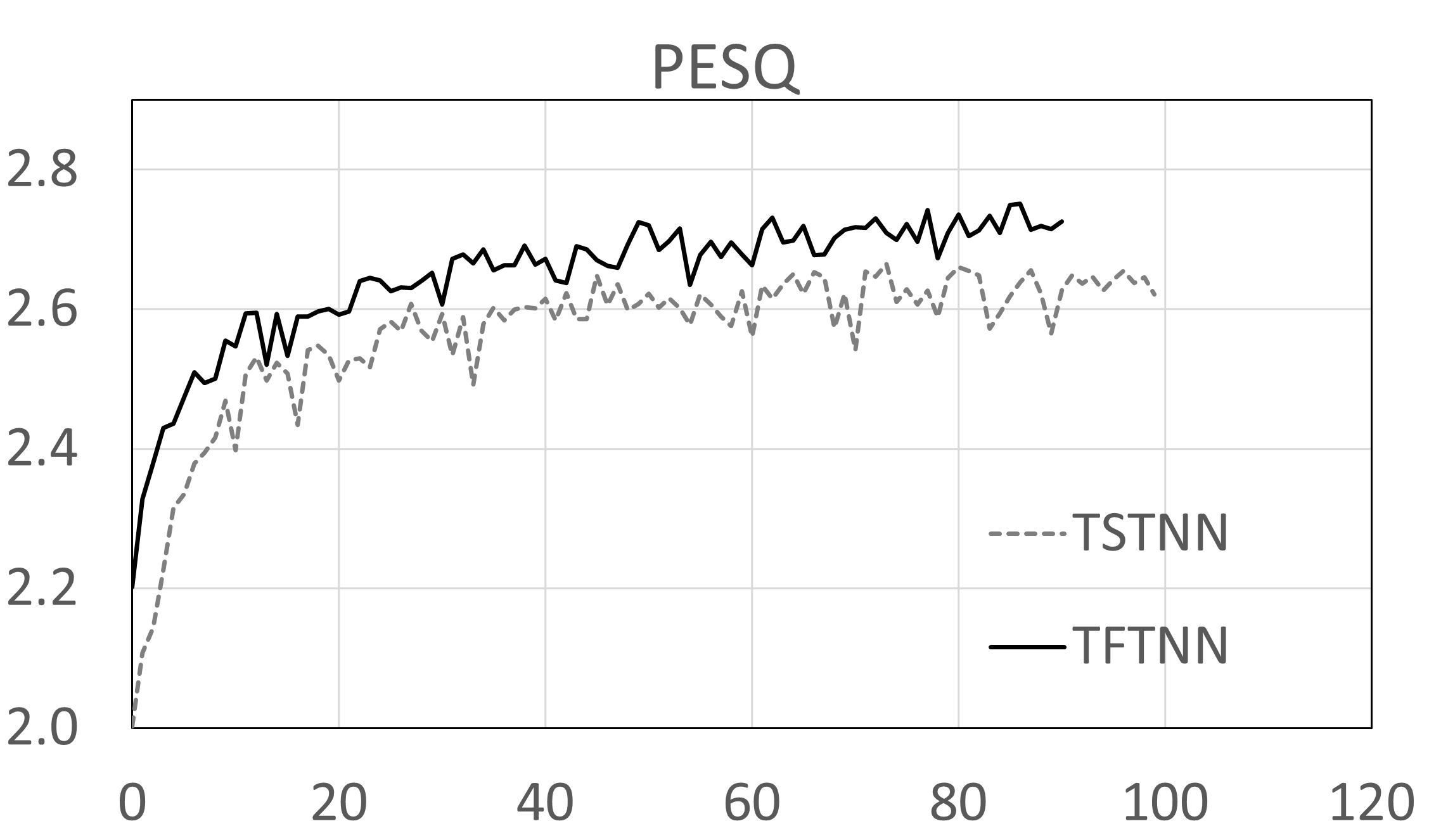}
\caption{Training curve of TSTNN (TF mask and spectrum loss) and TFTNN.}
\label{TFTNN_training}
\end{figure}

Our model is developed using PyTorch. TFTNN is trained using the Adam optimizer over 125 epochs. A batch size of 4 is selected, accounting for the batch-independent characteristics of the speech signal. As for the learning rate, it starts at 1e-3 and undergoes decay through the ReduceLROnPlateau function by a factor of 0.5, which adjusts the learning rate downward when performance plateaus.
Lastly, our loss function amalgamates both the time domain and spectrum domain losses, as depicted in Eq.~\ref{eq:loss_function}, with a value of $\alpha$ fixed at 0.2 for our experiments. Fig.~\ref{TFTNN_training} shows the training curve of the proposed TFTNN, which has a convergence curve similar to that of the TSTNN.

\begin{equation}
    \label{eq:loss_function}
    loss = \alpha \times loss_F + (1- \alpha) \times loss_T 
\end{equation}

\subsection{Model Performance and Comparisons}
As illustrated in Table~\ref{table:comparison with state of art}, the performance of TFTNN is benchmarked against other state-of-the-art models from recent literature. Computation metrics presented are derived from a 1-second data duration. Notably, our model's performance ranks just behind DBTNet\cite{DBT_Net}, albeit at a significantly reduced complexity of 40$\times$. Moreover, TFTNN outperforms TSTNN, attributed to the incorporated performance enhancement techniques, while achieving a 19.6$\times$ reduction in complexity. Hence, TFTNN demonstrates a competitive edge, combining robust performance with a highly streamlined model and efficient computational footprint.

\subsection{Ablation Study}
Owing to space constraints, a comprehensive listing of all results stemming from our analytical approaches is not feasible. To streamline the presentation, we emphasize the pivotal decisions below. Our simulations indicate that other methods exert minimal influence on performance.

\textit{Domain of the Mask and Loss Function:} Results from various domains for the mask and loss function are presented in Table~\ref{table:comparison of the different domain methods}. The data reveals a notable degradation in performance due to model compression, especially concerning the time-frequency mask and spectrum loss (PESQ drops from 2.6788 in TSTNN to 2.1190 in TFTNN with a TF mask and F loss). However, by incorporating cross-domain masking and loss, we can enhance the model's performance, closely aligning it with that of TSTNN.

\textit{Variations in Transformer Block Numbers:} Table~\ref{table:comparison of the different transformer numbers} displays outcomes based on the varying number of transformer blocks in TFTNN. Interestingly, while the performance remains consistent with more than two transformer blocks, utilizing 2 blocks surpasses the performance achieved with 3. We theorize that an even count of transformer blocks offers better balance, given the dual-stage processing inherent to each block. To strike an balance between the number of transformer blocks and optimal performance, we opted for 2 transformer blocks.

\textit{BN versus LN:} As detailed in Table~\ref{table:comparison of Transfomer with Batch Normalization and Layer Normalization}, there's a marked difference in performance between models using BN and those using LN. While models employing BN exhibit a decline in performance, the degradation is mitigated by the application of cross-domain masking and loss. Incorporating an additional BN within the MHA module further narrows the performance gap, bringing it close to the original benchmarks.

\textit{Quantization Considerations:}
In TFTNN, the feature maps display a broad dynamic range, spanning from $10^{-8}$ to 30 in absolute values, primarily attributed to the intricacies of speech-related processing. To accommodate this range, we experimented with various quantization schemes, as delineated in Table~\ref{table:Floating point quantization}, ensuring consistent bit allocation for both activation and weight. Notably, fixed-point quantization leads to a marked degradation when using less than 16 bits. Conversely, the losses are relatively insignificant with floating-point representations, credited to their expansive dynamic range. Consequently, we settled on the FP10 format (sign: 1, exponential: 5, mantissa: 4), striking a balance between bit-width and performance for our processing elements (PEs).

\begin{table}[htbp]
\caption{Quantization result of TFTNN on VoiceBank+UrbanSound8K.}
\label{table:Floating point quantization}
\begin{center}

\begin{tabular}{ccccccccc}
\hline
\multicolumn{1}{l}{\multirow{2}{*}{}} 
& \multicolumn{5}{c}{\multirow{2}{*}{\textbf{Bit}}} & \multicolumn{3}{c}{\multirow{2}{*}{\textbf{Post Quantization}}}  \\

\multicolumn{1}{l}{} 
& \multicolumn{5}{c}{} & \multicolumn{3}{c}{} \\ \hline

\multicolumn{1}{l}{} 
& \textbf{Act.} & \textbf{W.} & \multicolumn{1}{l}{\textbf{S}} & \textbf{Exp.} & \textbf{Man.} & \textbf{PESQ}   & \textbf{STOI} & \textbf{SNR} \\ \hline

\multirow{6}{*}{FP}  & 32 & 32 & 1 & 8 & 23 & 2.7459 & 0.8779 & 14.7450 \\ \cline{2-9} 
& 16  & 16   & \multirow{4}{*}{1}  & 8 & 7  & 2.7500 & 0.8778 & 14.7280    \\
& 10  & 10   &  & 5  & 4 & 2.7215  & 0.8760   & 13.0410    \\
& 9   & 9    &  & 4  & 4 & 2.6653  & 0.8714   & 11.7771    \\
& 8   & 8    &  & 4  & 3 & 2.5331  & 0.8596   & 9.3176     \\ \hline
                                      
\multicolumn{1}{l}{}                  & \textbf{Act.} & \textbf{W.} & \multicolumn{1}{l}{\textbf{S}} & \textbf{Int.} & \textbf{Dec.} & \textbf{PESQ}        & \textbf{STOI}       & \textbf{SNR}       \\ \hline
\multirow{4}{*}{FxP}          
& 16  & 16  & \multirow{4}{*}{1} & 8 & 7 & 2.7453 & 0.8774 & 14.7365 \\
& 10  & 10  &  & 5 & 4  & 2.2562 & 0.8469 & 6.7709 \\
& 9   & 9   &  & 4 & 4  & 1.9820 & 0.8266 & 3.7512 \\
& 8   & 8   &  & 4 & 3  & 1.7416 & 0.8033 & 3.5201 \\ \hline
\end{tabular}
\end{center}

\end{table}

\textit{Model size compression:}
Table~\ref{table:light weight model comparison} outlines the contributions of the primary four compression techniques, with the halving of channels making the most significant impact. The half-channel technique also has the largest performance impact. As shown in Table~\ref{table:comparison of the different domain methods}, the PESQ of the original TSTNN model is 2.639. With the cross-domain masking and loss, PESQ is increased to 2.8848. Then, with half channel and softmax-free attention, PESQ is drop to 2.7966 and 2.7658, respectively. However, it is still higher than the original one. Other techniques have smaller performance impact. In summary, the proposed techniques effectively compress the model, and their performance impact is unaffected. 

\begin{table}[htbp]
\begin{center}
\caption{Model reduction with the four main compression methods.}
\label{table:light weight model comparison}

\begin{tabular}{ccccccc}
\hline
\textbf{Model}   
& \textbf{\footnotemark[1]R.} 
& \textbf{\footnotemark[2]S.} 
& \textbf{\footnotemark[3]1/2 ch.}
& \textbf{\footnotemark[4]1/2 Tr.}
& \textbf{Size (K)} 
& \textbf{\footnotemark[5] (GMAC)} \\ \hline
TSTNN & \multicolumn{1}{l}{} & \multicolumn{1}{l}{}                    
& \multicolumn{1}{l}{}   &     & 922.87      & 9.87    \\ \hline
\multirow{4}{*}{\begin{tabular}[c]{@{}c@{}}TFTNN\\ \end{tabular}  } & \checkmark   &    &   &    & 449.95  & 3.83    \\
& \checkmark  & \checkmark   &  &   & 348.58  & 3.01   \\
& \checkmark   & \checkmark   & \checkmark   &   & 89.30     & 0.782    \\
& \checkmark  & \checkmark  & \checkmark  & \checkmark & 55.92  & 0.496    \\ \hline
\end{tabular}

\end{center}
\footnotemark[1]{Residual block with channel splitting.}
\footnotemark[2]{Subband attention only by removing full-band multi-head attention.}
\footnotemark[3]{half channels.}
\footnotemark[4]{Reduce transformer blocks}
\footnotemark[5]{1 second signal with 8K sampling rate.}

\end{table}

\subsection{Hardware Implementation Result}
Our proposed design was designed using Verilog and subsequently realized utilizing TSMC's 40nm CMOS technology. The design occupies an area of 0.367 $mm^2$ and integrates an SRAM of 53.75KB. With an operating frequency set at 62.5MHz, the core's power consumption is measured at 8.08 mW. Notably, this accelerator is equipped to execute TFTNN operations in real time, processing 512 samples (equivalent to 64 ms at an 8K sample rate) for every hop length, which spans 16 ms.

\subsubsection{Power analysis}
\begin{figure}[htbp]
\centering
\includegraphics[height=!,width=1.0\linewidth,keepaspectratio=true]{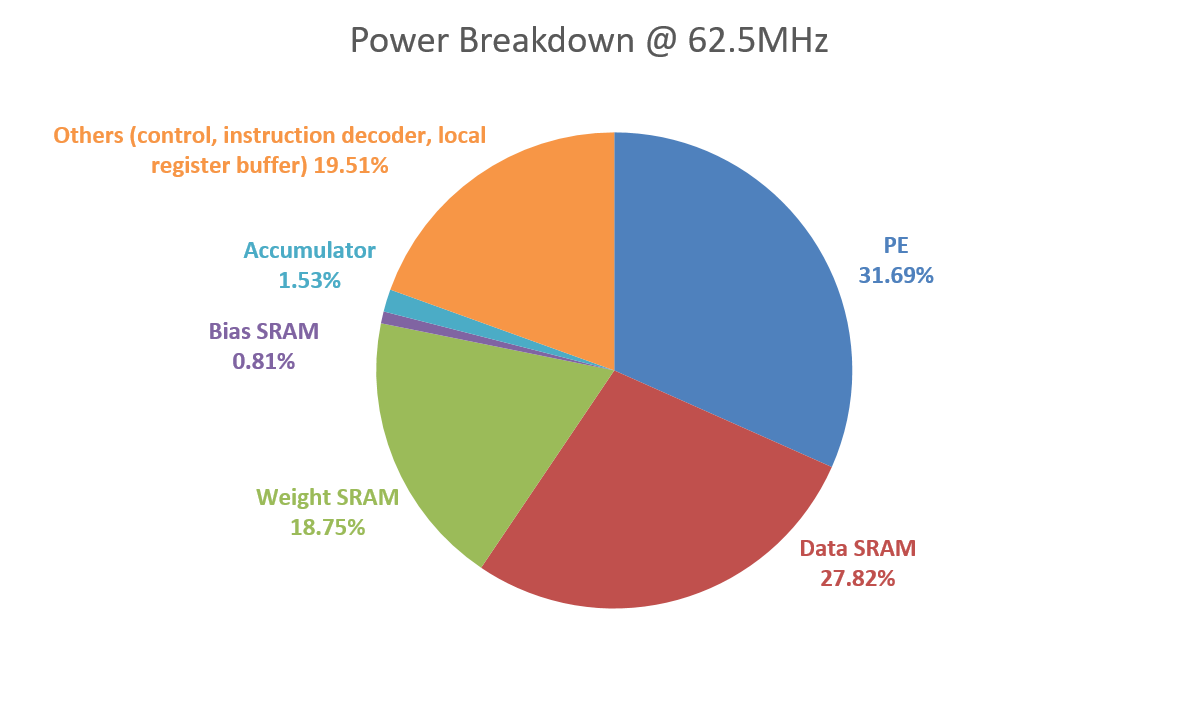}
\caption{The power breakdown of the core modules.}
\label{power_breakdown_module}
\end{figure}

Figure \ref{power_breakdown_module} provides a detailed breakdown of power consumption for each module within our design. The PE, data SRAM, and weight SRAM are responsible for 31.69\%, 27.82\%, and 18.75\% of the overall power consumption, respectively. We utilized Synopsys PrimeTime PX to simulate power consumption, basing our calculations on test speech data. In an effort to minimize the design's power draw, we implemented clock gating for idle SRAM banks, achieving a 5.4\% power savings for SRAM. Additionally, by leveraging ReLU for sparse input processing, we incorporated zero skipping to bypass computations involving zero values. This, combined with data and clock gating for the PEs, resulted in a substantial power reduction of 39.2\%.

\subsubsection{Design comparison}

For model compression, compared to the best previous model compression methods used for speech enhancement~\cite{tan2021towards} (small performance loss and 21X compression ratio with hardware unfriendly unstructured pruning and clustering-based quantization), our approach can achieve a 52.5X compression ratio (16.5X by hardware friendly structured pruning and 3.2X by quantization) without performance loss. This shows the benefits of adopting the co-design of model compression and target application.

Comparing our work with other designs is challenging due to variations in target applications and model architectures. For reference, Table~\ref{table:hardware_result} enumerates selected speech-related studies focused on recognition and enhancement, deliberately omitting works that are exclusively transformer-based and cater to generic vision or NLP tasks. We also normalize the area and power of other designs with the same process node and voltage in our design for a fair comparison. Compared to other designs, our design achieves the lowest area cost, 0.367$mm^2$, and the best normalized area efficiency, 21.798 GOPS/$mm^2$ when operating at the clock frequency of 250 MHz. This smallest area is due to several factors. First, due to our adoption of streaming inference, the design requires only 53.75 KB of SRAM, the lowest among the references cited. Furthermore, our approach uses a compact model, minimal PE numbers, and straightforward data flow, resulting in the smallest footprint and the best area efficiency compared to other designs. In particular, despite its compactness, our design remains versatile, accommodating a range of layer structures. With the smallest PE numbers, our design did not have the highest throughput and the corresponding energy efficiency, since our throughput target is to meet the required real-time constraints. Existing transformer-based designs only optimize transformer attention execution by exploiting the sparsity of attention~\cite{HPCA2020_A3, ISSCC2022_28nm, acceltran, vitcod} instead of the whole model as in this work. In addition, our design must optimize for CNN, transformer, and GRU at the same time, which is not addressed in previous designs.

\section{Conclusion}
\label{chapter:conclusion}

In this study, we introduce a 55.92K transformer-based model and its 8.08mW low-power design, specifically tailored for real-time streaming speech enhancement in edge devices. Leveraging model and hardware optimizations, we achieved a remarkable 93.9\% reduction in model size and a 94.9\% decrease in complexity. This was accomplished through the use of \textit{domain-aware and streaming-aware pruning}, all while preserving optimal performance via \textit{cross-domain masking and loss}. To enhance hardware compatibility, we transitioned to a BN-based approach, sidelining the traditional LN-based transformers. Additionally, we incorporated \textit{softmax-free attention complemented by an extra BN} to minimize latency. Our streamlined model enables our design to accommodate a diverse range of computing patterns, facilitated by a reconfigurable element-wise 1-D MAC array. Realized using the TSMC 40nm CMOS process, our design is both compact and efficient, encompassing only 207.8K gates and 53.75KB SRAM, with an energy footprint of just 8.08 mW for real-time tasks. With these programmable hardware designs, our design can be easily programmed to support a myriad of speech processing models in future applications.

\bibliographystyle{IEEEtran}

\bibliography{IEEEabrv, bib/thesis}

\begin{IEEEbiography}[{\includegraphics[width=1in,height=1.25in,clip,keepaspectratio]{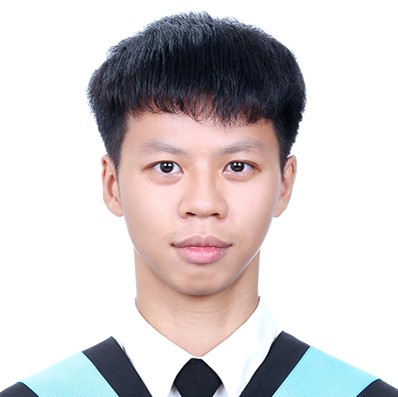}}]{Ci-Hao Wu}
received the M.S. degree in electronics engineering from the National Yang Ming Chiao Tung University, Hsinchu, Taiwan, in 2022. He is currently working in the Mediatek, Hsinchu, Taiwan. His research interest includes speech enhancement neural network design and VLSI design.

\end{IEEEbiography}

\begin{IEEEbiography}[{\includegraphics[width=1in,height=1.25in,clip,keepaspectratio]{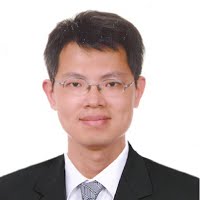}}]{Tian-Sheuan Chang}
	(S’93–M’06–SM’07)
	received the B.S., M.S., and Ph.D. degrees in electronic engineering from National Chiao-Tung University (NCTU), Hsinchu, Taiwan, in 1993, 1995, and 1999, respectively. 
	
	From 2000 to 2004, he was a Deputy Manager with Global Unichip Corporation, Hsinchu, Taiwan. In 2004, he joined the Department of Electronics Engineering, NCTU (as National Yang Ming Chiao Tung University (NYCU) in 2021), where he is currently a Professor. In 2009, he was a visiting scholar in IMEC, Belgium. His current research interests include system-on-a-chip design, VLSI signal processing, and computer architecture.
	
	Dr. Chang has received the Excellent Young Electrical Engineer from Chinese Institute of Electrical Engineering in 2007, and the Outstanding Young Scholar from Taiwan IC Design Society in 2010. He has been actively involved in many international conferences as an organizing committee or technical program committee member.
\end{IEEEbiography}
\end{document}